\documentclass[aps,prd,superscriptaddress,showpacs,preprintnumbers]{revtex4}
\usepackage{graphicx,color}
\newcommand{\be}{\begin{equation}}
\newcommand{\ee}{\end{equation}}
\newcommand{\bea}{\begin{eqnarray}}
\newcommand{\eea}{\end{eqnarray}}

\newcommand{\bfk}{\mbox{\boldmath $k$}}

\def\kt{k_\perp}
\newcommand{\bfp}{\mbox{\boldmath $p$}}

\newcommand{\bfP}{\mbox{\boldmath $P$}}
\newcommand{\bfS}{\mbox{\boldmath $S$}}

\newcommand{\pup}{p^\uparrow}

\def\xb{x_{_{\!B}}}

\def\lsim{\mathrel{\rlap{\lower4pt\hbox{\hskip1pt$\sim$}}\raise1pt\hbox{$<$}}}
\def\gsim{\mathrel{\rlap{\lower4pt\hbox{\hskip1pt$\sim$}}\raise1pt\hbox{$>$}}}
\def\nostrocostruttino#1\over#2{\mathrel{\mathop{\kern 0pt \rlap
{\hbox{$#1$}}} \hbox{\kern-.135em $#2$}}}

\begin{document}
\title{Sivers Effect for Pion and Kaon Production in \\ Semi-Inclusive Deep
Inelastic Scattering}
\author{M.~Anselmino}
\affiliation{Dipartimento di Fisica Teorica, Universit\`a di Torino,
             Via P. Giuria 1, I-10125 Torino, Italy}
\affiliation{INFN, Sezione di Torino, Via P. Giuria 1, I-10125 Torino, Italy}
\author{M.~Boglione}
\affiliation{Dipartimento di Fisica Teorica, Universit\`a di Torino,
             Via P. Giuria 1, I-10125 Torino, Italy}
\affiliation{INFN, Sezione di Torino, Via P. Giuria 1, I-10125 Torino, Italy}
\author{U.~D'Alesio}
\affiliation{Dipartimento di Fisica, Universit\`a di Cagliari,
             I-09042 Monserrato (CA), Italy}
\affiliation{INFN, Sezione di Cagliari,
             C.P. 170, I-09042 Monserrato (CA), Italy}
\author{A.~Kotzinian}
\affiliation{CEA-Saclay, IRFU/Service de Physique Nucl\'{e}aire, 91191 Gif-sur-Yvette, France}
\affiliation{Yerevan Physics Institute, 375036 Yerevan, Armenia} 	
\affiliation{JINR, 141980 Dubna, Russia}
\author{S.~Melis}
\affiliation{Dipartimento di Fisica Teorica, Universit\`a di Torino,
             Via P. Giuria 1, I-10125 Torino, Italy}
\affiliation{INFN, Sezione di Torino, Via P. Giuria 1, I-10125 Torino, Italy}
\author{F.~Murgia}
\affiliation{INFN, Sezione di Cagliari,
             C.P. 170, I-09042 Monserrato (CA), Italy}
\author{A.~Prokudin}
\affiliation{Di.S.T.A., Universit\`a del Piemonte Orientale
             ``A. Avogadro'', Alessandria, Italy}
\affiliation{Dipartimento di Fisica Teorica, Universit\`a di Torino,
             Via P. Giuria 1, I-10125 Torino, Italy}
\affiliation{INFN, Sezione di Torino, Via P. Giuria 1, I-10125 Torino, Italy}
\author{C.~T\"{u}rk}
\affiliation{Dipartimento di Fisica Teorica, Universit\`a di Torino,
             Via P. Giuria 1, I-10125 Torino, Italy}
\affiliation{INFN, Sezione di Torino, Via P. Giuria 1, I-10125 Torino, Italy}
\date{\today}

\begin{abstract}
We study the Sivers effect in the transverse single spin asymmetries (SSA) for
pion and kaon production in semi-inclusive deep inelastic scattering (SIDIS)
processes. We perform a fit of $A^{\sin(\phi_h-\phi_S)}_{UT}$ which, by
including recent high statistics experimental data for pion and kaon
production from HERMES and COMPASS Collaborations, allows a new determination
of the Sivers distribution functions for quarks and antiquarks with $u$, $d$
and $s$ flavours. Estimates for forthcoming SIDIS experiments at COMPASS
and JLab are given.
\end{abstract}


\maketitle

\section{\label{Intro} Introduction}

In Refs.~\cite{Anselmino:2005nn,Anselmino:2005ea}, we studied the
transverse single spin asymmetry $A_{UT}^{\sin(\phi_{h}-\phi_S)}$ observed
by the HERMES~\cite{Airapetian:2004tw} and COMPASS~\cite{Alexakhin:2005iw}
Collaborations in polarized SIDIS processes,
$\ell \, p\,(\bfS) \to \ell ^\prime \, h \, X$. The quality and amount of
the data allowed to perform a rather well constrained extraction of the
Sivers distribution function \cite{Sivers:1989cc,Sivers:1990fh} for $u$ and
$d$ quarks, assuming the existence of a symmetric and negligibly small
Sivers sea. Similar analyses and extractions were performed by other groups
\cite{Vogelsang:2005cs, Collins:2005ie, Anselmino:2005an}.
Although all these results were relevant and significant as the first
determination of the Sivers $u$ and $d$ functions, they were affected
by the low statistics of the experimental data available at that time:
in fact, COMPASS asymmetries were limited to charged hadron production,
as no hadron separation was performed, while the HERMES data were only
given for charged pion production. Recently, much higher statistics data
on the $A_{UT}^{\sin(\phi_{h}-\phi_S)}$ azimuthal asymmetries for SIDIS have
become available: in Ref.~\cite{Diefenthaler:2007rj} the HERMES Collaboration
presents neutral pion and charged kaon azimuthal asymmetries, in addition
to higher precision data on charged pion asymmetries; moreover,
Refs.~\cite{Martin:2007au,Alekseev:2008dn} show the COMPASS Collaboration
measurements for separated charged pion and kaon asymmetries,
together with some data for $K^0_S$ production.

It is then timely and natural to reconsider the analysis performed in
Ref.~\cite{Anselmino:2005ea} in order to increase our understanding of the
properties of the Sivers function. In particular, reduced error bars and
hadron separation in both the HERMES and COMPASS sets of experimental data
allow a better determination of the $u$ and $d$ flavour Sivers
distribution functions and, most importantly, a first insight into the sea
and strange contributions to the Sivers functions, namely
$\Delta^N \! f_ {\bar u/\pup}$, $\Delta^N \! f_ {\bar d/\pup}$,
$\Delta^N \! f_ {s/\pup}$ and $\Delta^N \! f_ {\bar s/\pup}$.

Our strategy is the following. First we evaluate the impact of the new data
with respect to the old data sets: as we shall explain in Section \ref{fit},
using the same unpolarized fragmentation functions as in
Refs.~\cite{Anselmino:2005nn,Anselmino:2005ea} would give a high quality fit
as far as pion asymmetries are concerned, but fails to describe the kaon data.
Instead, the use of a different, more recent set of fragmentation
functions~\cite{deFlorian:2007aj}, based on a global analysis of pion and kaon
production, will prove to be crucial to reach a successful description
of pion and kaon data simultaneously. With a simple ansatz parameterization of
the Sivers functions, we will then perform a simultaneous fit of both HERMES
and COMPASS data sets on $A^{\sin(\phi_h-\phi_S)}_{UT}$ for pion ($\pi^\pm, \,
\pi^0$) and $K^\pm$ production. We do not include in our fit the COMPASS data
on $K^0_S$ \cite{Alekseev:2008dn} as the corresponding fragmentation functions
are not so well established and can be obtained from those for $K^\pm$ only
adopting further assumptions; we shall rather estimate
$A^{\sin(\phi_h-\phi_S)}_{UT}$ for $K^0_S$, and compare it with COMPASS data,
using the Sivers functions obtained by fitting all other data sets, and
assuming exact $SU(2)$ invariance to derive the quark fragmentation functions
into $K^0_S$.

The above procedure will allow to determine the valence and sea proton Sivers
functions, which will be used to provide estimates for the analogous single
spin asymmetries that will soon be measured at JLab (operating on proton,
neutron and deuteron targets) and at COMPASS (operating on a proton target).
Notice that the JLab measurements will provide vital information on the large
$x$ behaviour of the Sivers distribution functions, yet undetermined from
present SIDIS experiments, as explained in Section~\ref{estimates}.

\section{\label{Mod} Formalism and parameterization}

\begin{figure}[t]
\begin{center}
\scalebox{0.26}{\input{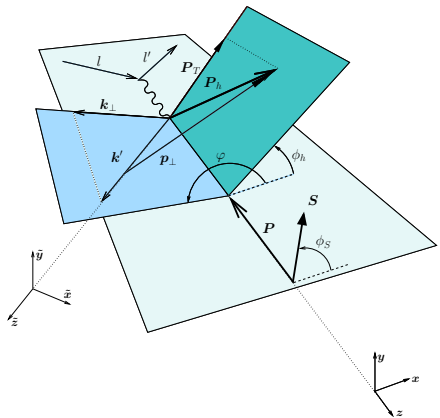}}
\caption{\small Kinematics of the SIDIS process in the $\gamma^* p$ center of
mass frame.}
\label{fig:planessidis}
\end{center}
\end{figure}

The SIDIS transverse single spin asymmetry (SSA) $A^{\sin(\phi_h-\phi_S)}_{UT}$
measured by HERMES and COMPASS is defined as (see Fig.~\ref{fig:planessidis}
for the definition of the azimuthal angles)
\be
A^{\sin (\phi_h-\phi_S)}_{UT} = \label{def-siv-asym}
2 \, \frac{\int d\phi_S \, d\phi_h \,
[d\sigma^\uparrow - d\sigma^\downarrow] \, \sin(\phi_h-\phi_S)}
{\int d\phi_S \, d\phi_h \,
[d\sigma^\uparrow + d\sigma^\downarrow]}\,,
\ee
and shows the azimuthal modulation triggered by the correlation between the
nucleon spin and the quark intrinsic transverse momentum. This effect is
embodied in the Sivers distribution function $\Delta^N \! f_ {q/\pup}(x,\kt)$,
which appears in the number density of unpolarized quarks $q$ with intrinsic
transverse momentum $\bfk _\perp$ inside a transversely polarized proton
$\pup$, with three-momentum $\bfP$ and spin polarization vector $\bfS$,
\bea
\hat f_ {q/\pup} (x,\bfk_\perp) &=& f_ {q/p} (x,\kt) +
\frac{1}{2} \, \Delta^N \! f_ {q/\pup}(x,\kt)  \;
{\bfS} \cdot (\hat {\bfP}  \times
\hat{\bfk}_\perp) \,,\label{sivnoi}
\eea
where $f_ {q/p}(x,\kt)$ is the unpolarized $x$ and $\kt$ dependent
parton distribution, and the mixed product
${\bfS} \cdot (\hat {\bfP}  \times \hat{\bfk}_\perp)$ explicitly gives
the azimuthal dependence mentioned above. Notice that the Sivers function
is also often denoted as $f_{1T}^{\perp q}(x, k_\perp)$ \cite{Mulders:1995dh};
this notation is related to ours by \cite{Bacchetta:2004jz}
\be
\Delta^N \! f_ {q/\pup}(x,k_\perp) = - \frac{2\,k_\perp}{m_p} \>
f_{1T}^{\perp q}(x, k_\perp) \>. \label{rel}
\ee

The ``weighting'' factor $\sin(\phi_h -\phi_S)$ in Eq.~(\ref{def-siv-asym})
is appropriately chosen to single out, among the various azimuthal dependent
terms appearing in $[d\sigma^\uparrow - d\sigma^\downarrow]$
\cite{Bacchetta:2006tn,SIDIS-general}, only the contribution of the Sivers
mechanism. By properly taking into account all intrinsic motions this
transverse single spin asymmetry can be written, at order $(\kt/Q)$,
as \cite{Anselmino:2005ea}
\be
A^{\sin (\phi_h-\phi_S)}_{UT} = \label{hermesut}
\frac{\displaystyle  \sum_q \int
{d\phi_S \, d\phi_h \, d^2 \bfk _\perp}\;
\Delta^N \! f_{q/\pup} (x,\kt) \sin (\varphi -\phi_S) \;
\frac{d \hat\sigma ^{\ell q\to \ell q}}{dQ^2} \;
\; D_q^h(z,p_\perp) \sin (\phi_h -\phi_S) }
{\displaystyle \sum_q \int {d\phi_S \,d\phi_h \, d^2 \bfk _\perp}\;
f_{q/p}(x,k _\perp) \; \frac{d \hat\sigma ^{\ell q\to \ell q}}{dQ^2} \;
 \; D_q^h(z,p_\perp) } \> \cdot
\ee
$\phi_S$ and $\phi_h$ are the azimuthal angles identifying the
directions of the proton spin $\bfS$ and of the outgoing hadron $h$
respectively, while $\varphi$ defines the direction of the incoming
(and outgoing) quark transverse momentum,
$\bfk_\perp$ = $\kt(\cos\varphi, \sin\varphi,0)$,
as shown in Fig.~\ref{fig:planessidis};
$\frac{d \hat\sigma ^{\ell q\to \ell q}}{dQ^2}$ is the unpolarized
cross section for the elementary scattering  $\ell q\to \ell q$,
\be
\frac{d \hat\sigma^{\ell q\to \ell q}}{d Q^2} = e_q^2 \,
\frac{2\pi \alpha^2}{\hat s^2}\,
\frac{\hat s^2+\hat u^2}{Q^4}\;,
\label{part-Xsec}
\ee
where $\hat s$, $\hat t = -Q^2$ and $\hat u$ are the partonic Mandelstam
invariants.

Finally, $D_q^h(z,p_\perp)$ is the fragmentation function describing the
hadronization of the final quark $q$ into the detected hadron $h$ with
momentum $\bfP_h$ (see Fig.~\ref{fig:planessidis}); $h$ carries, with respect
to the fragmenting quark, a light-cone momentum fraction $z$ and a transverse
momentum $\bfp_\perp$.
%
%

In our analysis we shall consider $u$, $d$ and $s$ flavours for quarks and
antiquarks. The Sivers function is parameterized in terms of the unpolarized
distribution function, as in Ref.~\cite{Anselmino:2005ea}, in the following
factorized form:
\be
\Delta^N \! f_ {q/\pup}(x,\kt) = 2 \, {\cal N}_q(x) \, h(\kt) \,
f_ {q/p} (x,\kt)\; , \label{sivfac}
\ee
with
\bea
&&{\cal N}_q(x) =  N_q \, x^{\alpha_q}(1-x)^{\beta_q} \,
\frac{(\alpha_q+\beta_q)^{(\alpha_q+\beta_q)}}
{\alpha_q^{\alpha_q} \beta_q^{\beta_q}}\; ,
\label{siversx} \\
&&h(\kt) = \sqrt{2e}\,\frac{k_\perp}{M_{1}}\,e^{-{k_\perp^2}/{M_{1}^2}}\; ,
\label{siverskt}
\eea
where $N_q$, $\alpha_q$, $\beta_q$ and $M_1$ (GeV/$c$) are free parameters
to be determined by fitting the experimental data. Since $h(\kt) \le 1$ for
any $\kt$ and $|{\cal N}_q(x)| \le 1$ for any $x$ (notice that we allow the
constant parameter $N_q$ to vary only inside the range $[-1,1]$), the
positivity bound for the Sivers function,
\be
\frac{|\Delta^N\!f_ {q/\pup}(x,\kt)|}{2 f_ {q/p} (x,\kt)}\le 1\>,
\label{pos}
\ee
is automatically fulfilled. We adopt the usual (and convenient) Gaussian
factorization for the unpolarized distribution and fragmentation functions:
\be
f_{q/p}(x,k_\perp) = f_q(x) \, \frac{1}{\pi \langle\kt^2\rangle} \,
e^{-{\kt^2}/{\langle\kt^2\rangle}}
\label{partond}
\ee
and
\be
D_q^h(z,p _\perp) = D_q^h(z) \, \frac{1}{\pi \langle p_\perp^2\rangle}
\, e^{-p_\perp^2/\langle p_\perp^2\rangle} \>,
\label{partonf}
\ee
with the values of $\langle k_\perp^2\rangle$ and $\langle p_\perp^2\rangle$
fixed to the values found in Ref.~\cite{Anselmino:2005nn} by analysing the
Cahn effect in unpolarized SIDIS:
\be
\langle\kt^2\rangle   = 0.25  \;({\rm GeV}/c)^2 \quad\quad\quad
\langle p_\perp^2\rangle  = 0.20 \;({\rm GeV}/c)^2 \>.
\label{ktpar}
\ee
Notice that the Gaussian distributions limit the effective action of
intrinsic motion to $\kt \lsim \sqrt{\langle\kt^2\rangle}$ and
$p_\perp \lsim \sqrt{\langle p_\perp^2\rangle}$, which is the region of
validity of the TMD factorized expressions in Eq.~(\ref{hermesut}),
$P_T\simeq \kt \simeq \Lambda _{\rm QCD} \ll
Q$~\cite{Ji:2004wu,Ji:2004xq,Ji:2006br}.

The parton distribution functions (PDF) $f_{q}(x)$ and the fragmentation
functions (FF) $D_q^h(z)$ also depend on $Q^2$ via the usual QCD evolution,
which will be taken into account, at leading order (LO), in all our computations.

\begin{figure}[t]
\vskip -24pt
\includegraphics[width=0.62 \textwidth,bb= 10 140 540 660,angle=-90]
{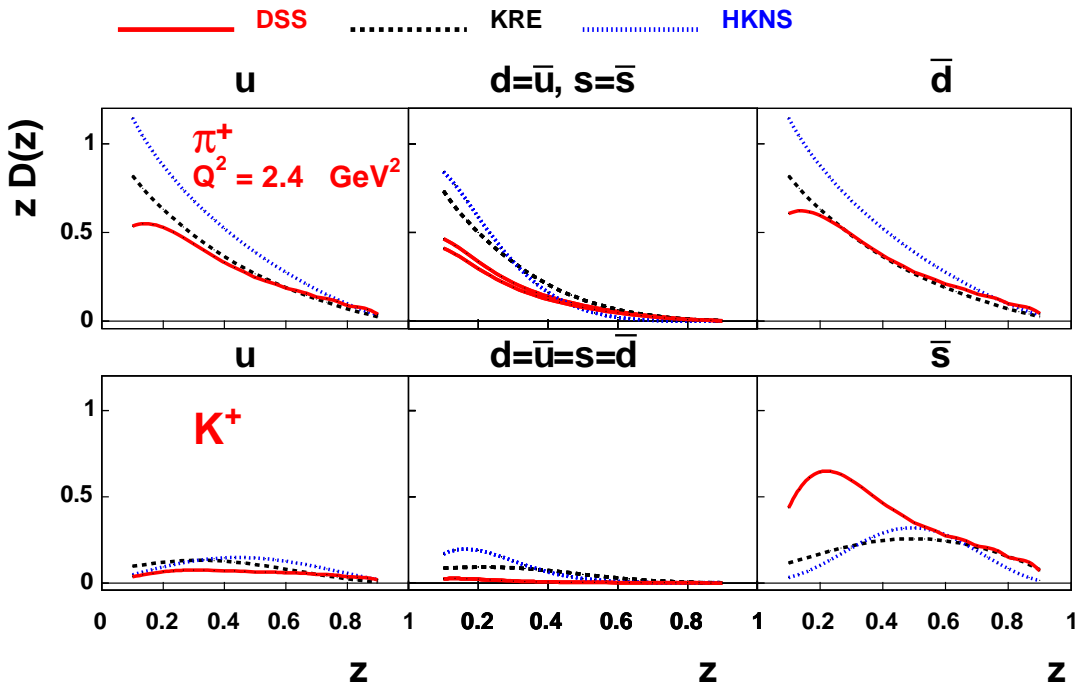}
\caption{\label{fig:ffs}
The LO unpolarized fragmentation functions for $u$, $d$ and $s$ flavours, at
$Q^2 = 2.4$ (GeV/$c)^2$, as given
by Kretzer \cite{Kretzer:2000yf} (dashed lines), by Hirai, Kumano, Nagai and
Sudoh \cite{Hirai:2007cx} (dotted lines) and by de Florian, Sassot and
Stratmann~\cite{deFlorian:2007aj} (solid lines). We show the fragmentation
functions for $\pi^+$ in the upper panel, and for $K^+$ production in the lower
panel. Notice that the fragmentation function describing the probability of a
$K^+$ originating from an $\bar s$ quark is much larger than that associated
to a $K^+$ from a $u$ quark, over the whole $z$ range. The two solid curves in
the central upper plot correspond to $zD_{d}^{\pi^+} = zD_{\bar u}^{\pi^+}$
(upper line) and $zD_{s}^{\pi^+} = zD_{\bar s}^{\pi+}$ (lower line).}
\end{figure}

Before fitting the data on the Sivers asymmetries a few comments on the quark
hadronization are necessary. While most of the available sets of fragmentation
functions describe rather well the pion multiplicities observed at HERMES,
many of them fail to reproduce the kaon multiplicities in SIDIS production.
The main reason is the role of the strange quarks, which is often not well
established: for example, one expects that $K^+$ mesons can be abundantly
produced by $\bar s$ quarks, via creation from the vacuum of a light $u \bar u$
pair, rather than by $u$ quarks, via creation from the vacuum of a heavier
$s \bar s$ pair. Such a feature is particularly emphasized in the set recently
obtained by de Florian, Sassot, Stratmann (DSS) \cite{deFlorian:2007aj},
which has $D_{\bar s}^{K^+}(z) \, \gg \, D_{u}^{K^+}(z)$ over the whole $z$
range. This is shown in Fig.~\ref{fig:ffs}, where the LO DSS fragmentation
functions (solid lines) are compared with those proposed by Kretzer
(KRE)~\cite{Kretzer:2000yf} (dashed lines) and by Hirai, Kumano, Nagai
and Sudoh (HKNS)~\cite{Hirai:2007cx} (dotted lines). The DSS set, which is
determined by fitting all presently available multiplicity measurements, both
for pions and kaons, is indeed the most suitable for our purposes.

This can also be seen in a more quantitative way. We know that Kretzer's and
other commonly adopted sets of fragmentation functions are able
to describe pion production data, as shown, for instance, in Fig.~4 of
Ref.~\cite{deFlorian:2007aj}. However, Fig.~13 of Ref.~\cite{deFlorian:2007aj}
shows instead that Kretzer fragmentation functions fail to reproduce charged
kaon SIDIS multiplicities, and might not be adequate to reconstruct transverse
single spin asymmetries corresponding to kaon production. In fact, by using
the Kretzer set for our fit, we would not be able to describe the kaon
asymmetry data: to be more precise, we would obtain $\chi ^2/d.o.f. \equiv
\chi^2_{dof} \simeq 1$ for pions but $\chi^2_{dof}\simeq 4$ for kaon
production asymmetries. Estimates for $K^\pm$ asymmetries were presented in
Ref. \cite{Anselmino:2005ea} and the inadequacy of the Kretzer fragmentation
functions was pointed out in several talks (see, for example, Ref. \cite{talk}).
The same conclusion has been confirmed, very recently, in
Ref. \cite{Arnold:2008ap}.

Let us now turn to the experimental data on kaon and pion Sivers azimuthal
asymmetries measured by the HERMES Collaboration~\cite{Diefenthaler:2007rj}.
The single spin asymmetry corresponding to $K^+$ production is, as a matter
of fact, much larger than the analogous asymmetry for $\pi^+$. Although one
could naively expect, on the basis of $u$ quark dominance, that $K^+$ and
$\pi^+$ asymmetries should be roughly the same, the presence of a large
$D_{\bar s}^{K^+}$ FF can help to understand the ``puzzle'' of the $K^+$
asymmetry. Indeed, if a non-negligible $\bar s$ Sivers function exists, then
its action combined with a large $D_{\bar s}^{K^+}$ fragmentation function
can give rise to a significant difference between $K^+$ and $\pi^+$ Sivers
asymmetries.
%
%

\section{Fit of SIDIS data and extraction of Sivers functions}\label{fit}

The recent SIDIS experimental data on Sivers asymmetries for pion
\textit{and kaon} production give us the opportunity to study sea-quark Sivers
functions for $\bar u$, $\bar d$, $s$ and $\bar s$ quarks. The $u$ and $d$ quark
Sivers functions alone were already studied in Ref. \cite{Anselmino:2005ea}; in
the present analysis we will be able to improve the extraction of these
functions and to present first estimates of the sea-quark Sivers functions.

In order to evaluate the significance of the sea-quark Sivers contributions we
first perform a fit of the SIDIS data using flavour independent ratios of the
sea-quark Sivers functions with the corresponding unpolarized PDFs: that is,
for $\bar u$, $\bar d$, $s$ and $\bar s$ flavours we attempt an ``unbroken sea"
ansatz:
\bea
&&\Delta^N \! f_ {q/\pup}(x,\kt) = 2 \, {\cal N}_{sea}(x) \, h(\kt) \,
f_ {q/p} (x,\kt)\; , \nonumber \\
&&{\cal N}_{sea}(x) =  N_{sea} \, x^{\alpha_{sea}}(1-x)^{\beta_{sea}}
\,\frac{(\alpha_{sea}+\beta_{sea})^{(\alpha_{sea}+\beta_{sea})}}
{\alpha_{sea}^{\alpha_{sea}} \beta_{sea}^{\beta_{sea}}}\; ,
\label{siverskt_sea}
\eea
where $N_{sea}, \alpha_{sea}, \beta_{sea}$ are the same for all sea quarks,
$q = \bar u$, $\bar d$, $s$ and $\bar s$.

As the SIDIS data from HERMES and COMPASS have a limited coverage in $x$,
typically $x < 0.3-0.4$, the experimental asymmetries we are fitting contain
very little information on the large $x$ tail of the Sivers functions.
In fact, our previous analysis of Ref.~\cite{Anselmino:2005ea} showed that
the parameters $\beta_{u}$ and $\beta_{d}$ as determined by MINUIT best fit
procedure are affected by very large errors. Therefore, as a first attempt,
we assume the same value of $\beta$ for all Sivers functions, setting
$\beta_{sea} = \beta_{u} = \beta_{d} \equiv \beta$.

Thus for the ``unbroken sea" ansatz we have 8 free parameters:
\bea
&& N_u \quad\quad\quad  N_d  \quad\quad\quad  N_{sea}
\nonumber \\
&& \alpha_u \quad\quad\quad \,\alpha_d \quad\quad\quad \;\alpha_{sea} \\
&& \beta   \quad\quad\quad\,\,\, M_1\;({\rm GeV}/c) \>. \nonumber
\label{par_unbroken}
\eea

For the purposes of our fit, we use the unpolarized parton distribution
functions $f_q(x,Q^2)$ as given in Ref.~\cite{Gluck:1998xa} (GRV98LO) and
the fragmentation functions $D_q^h(z,Q^2)$ as given in
Ref.~\cite{deFlorian:2007aj} (DSS) -- all evolved to the appropriate
$Q^2$ values -- with the additional $\kt$ Gaussian dependences of
Eqs.~(\ref{partond})-(\ref{ktpar}). While the choice of the DSS
fragmentation functions is the one which best describes the large
asymmetries observed for $K^+$, the use of different sets of distribution
functions, including the most recent analysis of $s$ quark distributions
from HERMES \cite{Airapetian:2008qf},
would not affect our results significantly.
For the Sivers functions,
we use the functional forms of Eqs.~(\ref{sivfac})-(\ref{siverskt}).
Notice that the (unknown) $Q^2$ evolution of these functions is assumed to be
the same as for the unpolarized PDFs, $f_q(x,Q^2)$.

By fitting simultaneously pion and kaon production data from
HERMES~\cite{Diefenthaler:2007rj} and COMPASS~\cite{Martin:2007au}
we obtain an acceptable overall description of the experimental data, with
$\chi^2_{dof} = 1.16$. The new experimental data give clear indications of the
need of a non negligible sea-quark Sivers function, with
$N_{sea} = -0.13 \pm 0.03$ sensitively different from zero.
Although the total $\chi^2_{dof}$ is definitely good, a more careful
examination of the results shows that while we achieve a perfect description
of the $\pi^+$ production data at HERMES~\cite{Diefenthaler:2007rj}, with
$\chi^2 \simeq 1$ per data point, the description of kaon production data is
rather poor, with $\chi^2 \simeq 3$ per data point for $K^+$ production at
HERMES~\cite{Diefenthaler:2007rj}. This indicates that the ``unbroken sea"
ansatz fails to reproduce the differences between pion and kaon production,
and clearly suggests the need of a parameterization which should allow for a
more structured flavour dependence of the sea-quark Sivers functions.

Including four new functions in our analysis would result in a substantial
growth of the number of parameters and would consequently limit
the usefulness of our parameterization.
To keep the number of parameters under control, we define a
simple ``broken sea" ansatz by introducing four free parameters,
$N_{\bar u}$, $N_{\bar d}$, $N_{s}$, and $N_{\bar s}$ which give different sizes
to the sea-quark Sivers functions, while keeping the same functional forms
($\alpha_{\bar u} = \alpha_{\bar d} = \alpha_{s} = \alpha_{\bar s} \equiv
\alpha_{sea}$ and $\beta_{sea} = \beta_{u} = \beta_{d} \equiv \beta$).
For the ``broken sea" ansatz fit we then have 11 parameters:
\bea
&& N_u \quad\quad\quad  N_d \quad\quad\quad  N_s
\nonumber \\
&& N_{\bar u} \quad\quad\quad  N_{\bar d} \quad\quad\quad  N_{\bar s}
\nonumber \\
&& \alpha_u \quad\quad\quad \,\alpha_d \quad\quad\quad \;\alpha_{sea} \label{par_broken}\\
&& \beta   \quad\quad\quad\,\,\, M_1\;({\rm GeV}/c) \>. \nonumber
\eea

The results we obtain for these parameters by fitting simultaneously the four
experimental data sets on $A_{UT}^{\sin(\phi_h-\phi_S)}$, corresponding to
pion and kaon production at HERMES~\cite{Diefenthaler:2007rj} and
COMPASS~\cite{Martin:2007au}, are presented in Table~\ref{fitpar_sivers},
together with the corresponding errors, estimated according to the procedure
outlined in Appendix \ref{stat}. The fit performed under the ``broken sea"
ansatz shows a remarkable improvement, especially concerning the description
of kaon data. We now obtain $\chi^2 = 1.20$ per data point for $K^+$
production at HERMES~\cite{Diefenthaler:2007rj}, while for pions we have
$\chi^2 = 0.94$ per data point, and a total $\chi^2_{dof} = 1.00$.
In Table II we show the $\chi^2$ per data point for pion and kaon production
at HERMES and COMPASS, both for the "unbroken sea" and "broken sea" ansatze
and adopting the Kretzer and DSS FF sets. Notice that these values refer
to the asymmetries as a function of $x$.

\begin{figure}[t]
\begin{center}
\includegraphics[width=0.35\textwidth,bb= 10 140 540 660,angle=-90]
{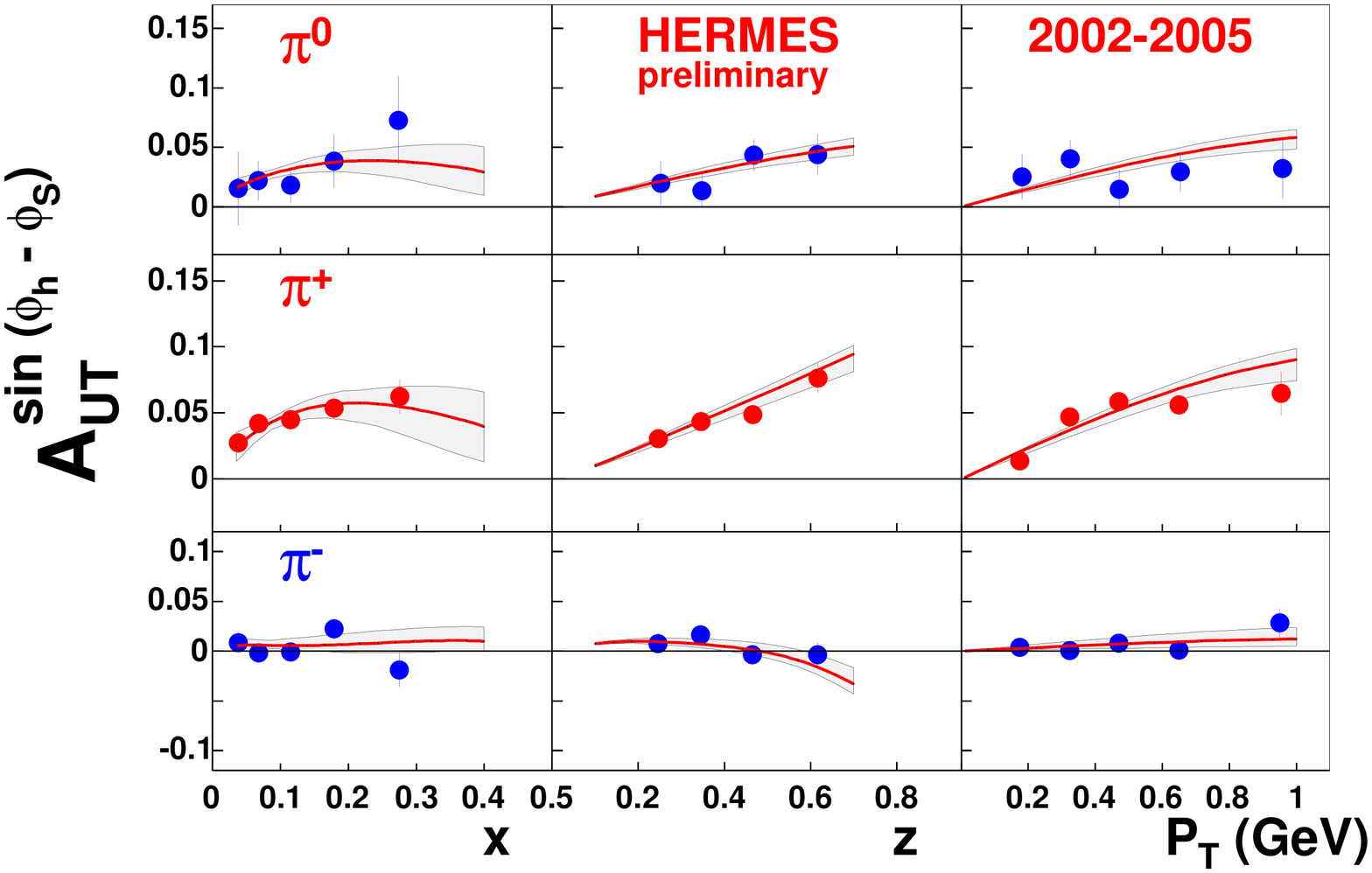} \hskip 2.85cm
\includegraphics[width=0.35\textwidth,bb= 10 140 540 660,angle=-90]
{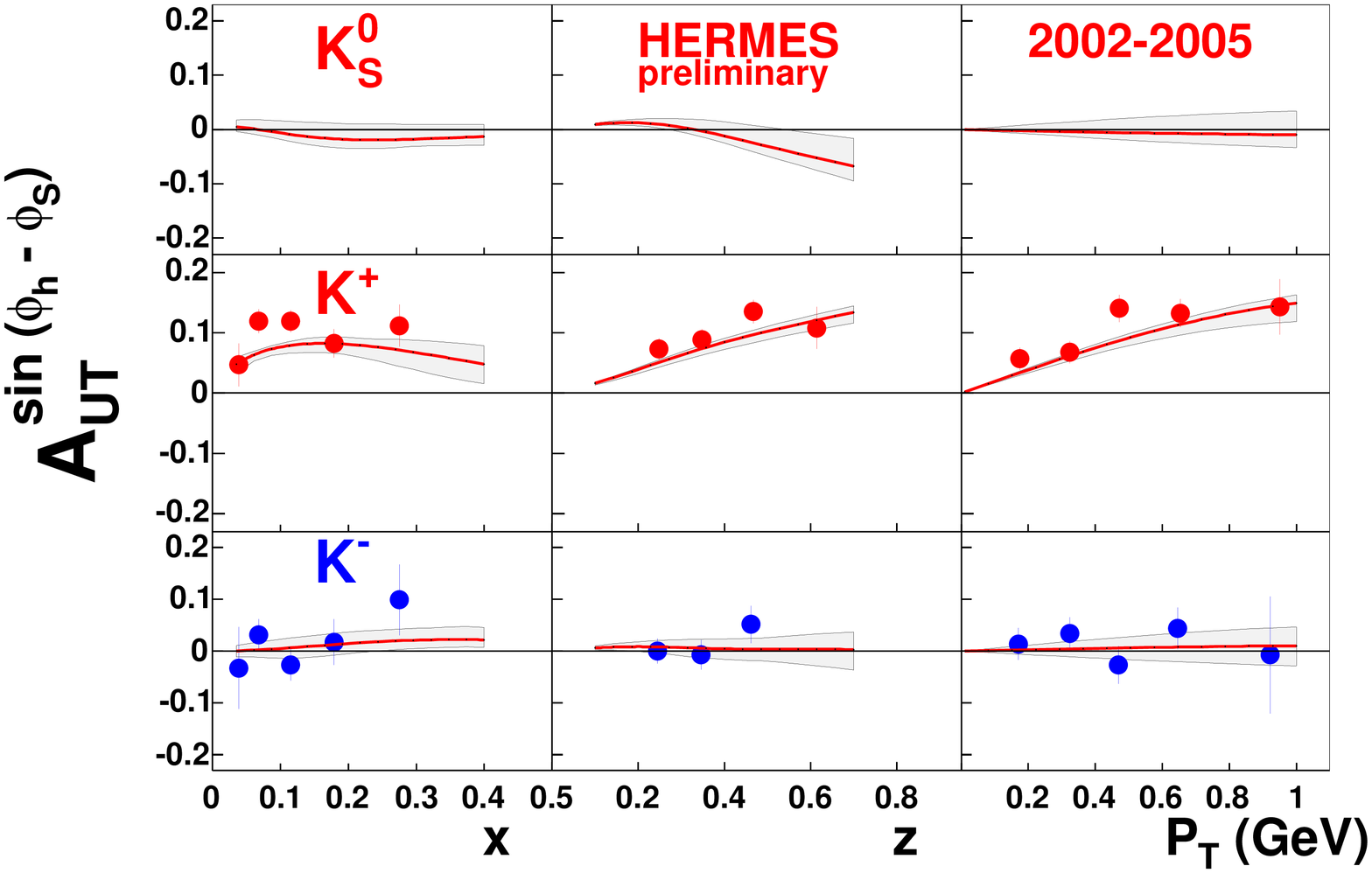}
\caption{\label{fig:hermes}
The results obtained from our simultaneous fit of the SIDIS
$A_{UT}^{\sin{(\phi_h-\phi_S)}}$ Sivers asymmetries (solid lines) are
compared with HERMES experimental data \cite{Diefenthaler:2007rj} for pion
and kaon production (left and right panel, respectively). The shaded area
corresponds to the statistical uncertainty of the parameters, see
Appendix~\ref{stat} for further details. For completeness, we also show the
$K_S^0$ asymmetry, not measured at HERMES, which is the result of a
computation based on our extracted Sivers function and the assumed
fragmentation functions of Eq.~(\ref{D0SFF}).}
\end{center}
\end{figure}

\begin{figure}[t]
\begin{center}
\includegraphics[width=0.35\textwidth,bb= 10 140 540 660,angle=-90]
{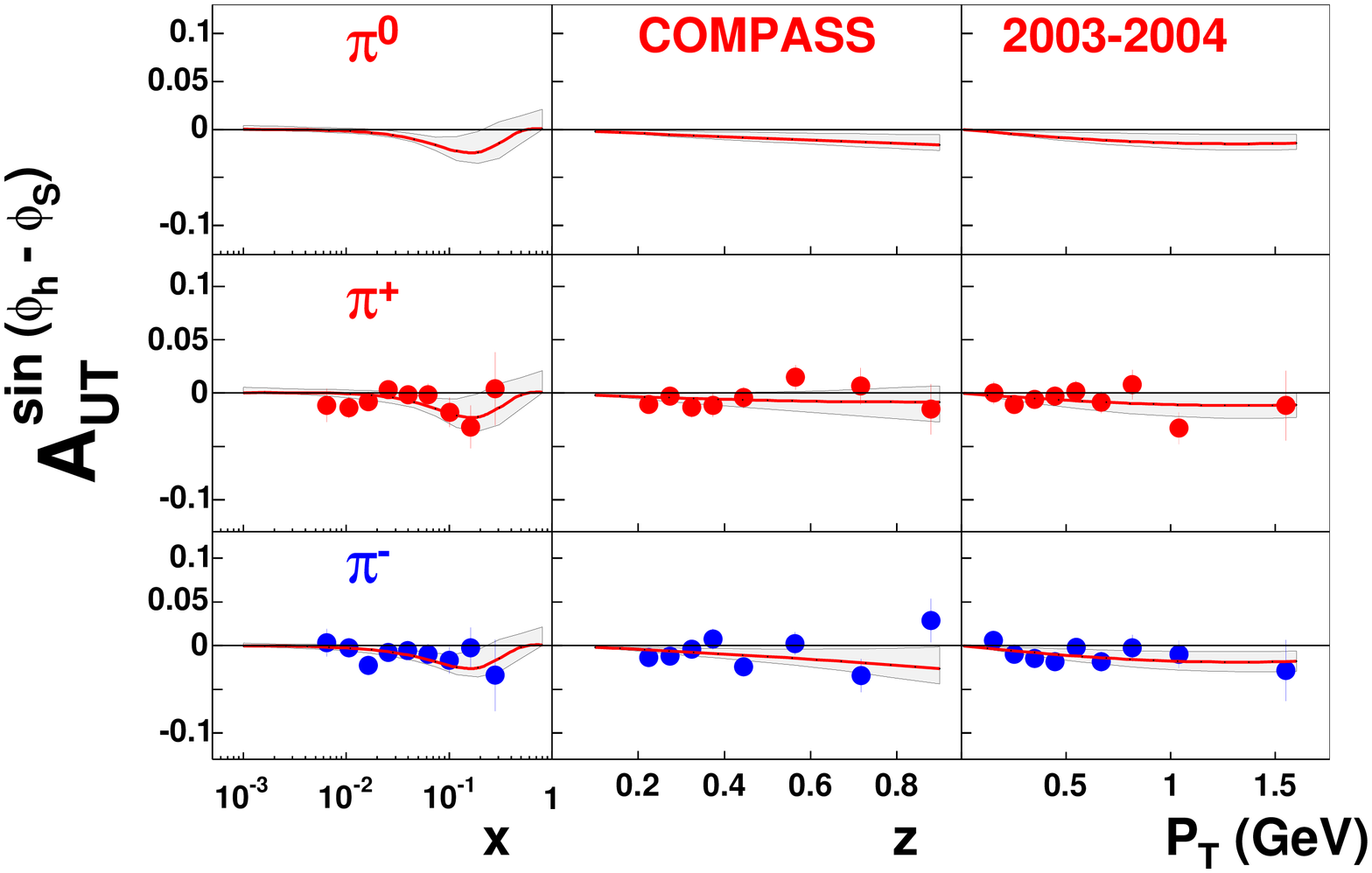} \hskip 2.85cm
\includegraphics[width=0.35\textwidth,bb= 10 140 540 660,angle=-90]
{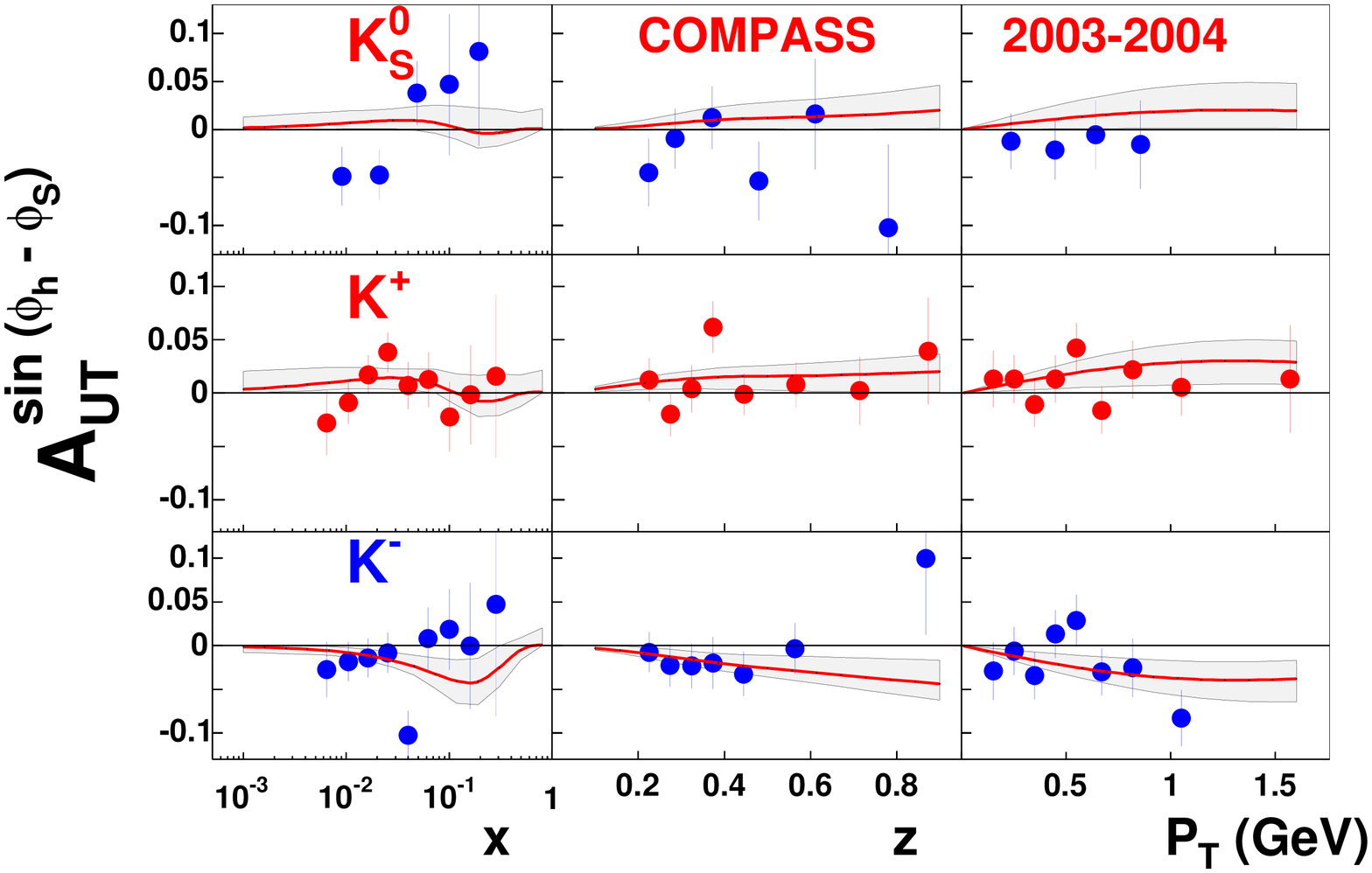}
\caption{\label{fig:compass}
The results obtained from our fit (solid lines) are compared with the COMPASS
measurements of $A_{UT}^{\sin{(\phi_h-\phi_S)}}$ for pion (left panel)
and kaon (right panel) production \cite{Martin:2007au} off a deuteron target.
The shaded area corresponds to the statistical uncertainty of the parameters,
as explained in Appendix~\ref{stat}. The $\pi^0$ asymmetry, not measured at
COMPASS, is the result of a computation based on our extracted Sivers functions.
Also the $K_S^0$ asymmetry, although compared with data~\cite{Alekseev:2008dn},
is not a best fit, but the result of our computation, using the assumed
fragmentation functions of Eq.~(\ref{D0SFF}).}
\end{center}
\end{figure}

The quality of our results is shown in Figs.~\ref{fig:hermes} and
\ref{fig:compass} where our best fit to the SSA is compared with the
experimental data from Refs.~\cite{Diefenthaler:2007rj}~and~\cite{Martin:2007au}:
the SSAs are plotted as a function of one variable at a time, either $x$ or $z$
or $P_T$, while an integration over the other variables has been performed
consistently with the cuts of the corresponding experiment.
%
%
\begin{table}[t]
\caption{
Best values of the free parameters for the ``broken sea" ansatz,
Eq.~(\ref{par_broken}).
Notice that the statistical errors reported in this table are not the errors given by
MINUIT. As the parameters are strongly correlated, we determine them
according to the procedure explained in Appendix A.
The significant fluctuations in our results are shown by the shaded areas in
Figs.~\ref{fig:hermes} and \ref{fig:compass}.
\label{fitpar_sivers}}
\begin{center}
\begin{tabular}{l l l}
\hline
\hline
\noalign{\vspace{3pt}}
\multicolumn{3}{ c }{~~~~~~~~~~~~~~~~~~$\chi^2/d.o.f. =
1.00$~~~~~~~~~~~~~~~~~~~~}\\
\noalign{\vspace{3pt}}
\hline
~&~&~\\
~~~$N_{u} = 0.35 ^{+ 0.08}_{-0.08} $ \hspace*{1cm} &
~~~$N_{d} = -0.90 ^{+ 0.43}_{-0.10} $ &
~~~$N_{s} = -0.24 ^{+ 0.62}_{-0.50} $~~ \\
~~~$N_{\bar u} =  0.04 ^{+ 0.22}_{-0.24} $&
~~~$N_{\bar d} =  -0.40 ^{+ 0.33}_{-0.44}$&
~~~$N_{\bar s} =  1 ^{+0}_{-0.0001}$ \\
~~~$\alpha _u = 0.73 ^{+ 0.72}_{-0.58}$ &
~~~$\alpha_d = 1.08 ^{+ 0.82}_{-0.65}$  &
~~~$\alpha_{sea} = 0.79 ^{+ 0.56}_{-0.47}$ \\
~~~$\beta \;\;= 3.46 ^{+ 4.87}_{-2.90}$ &
~~~$M_1^2 = 0.34 ^{+ 0.30}_{-0.16}$ (GeV/$c)^2$~  &
~~~~~ \\
~&~&~\\
\hline
\hline
\end{tabular}
\end{center}

\end{table}
%
\begin{table}[b]
\caption{$\chi^2$ per data point for pion and kaon production at HERMES and
COMPASS, both for the ``unbroken sea" and ``broken sea" ansatze, and adopting the
Kretzer and DSS FF sets. $x$ dependent data, integrated over $z$ and $P_T$,
are considered here.}
\vspace*{6pt}
\begin{ruledtabular}
\begin{tabular}{ccccccc}
\noalign{\vspace{3pt}}
  Experiment & observed hadron & n. of data points & \multicolumn{4}{c} {$\chi^2$ per data point} \\
\noalign{\vspace{3pt}}
\colrule
\noalign{\vspace{3pt}}
 & & & \multicolumn{2}{c}{Kretzer} & \multicolumn{2}{c}{DSS} \\
\noalign{\vspace{3pt}}
\cline{4-7}
\noalign{\vspace{3pt}}
 & & & \multicolumn{1}{c}{unbroken sea} &
 \multicolumn{1}{c}{broken sea} &  \multicolumn{1}{c}{unbroken sea} &
 \multicolumn{1}{c}{broken sea} \\
\noalign{\vspace{3pt}}
\cline{1-7}
\noalign{\vspace{3pt}}
          &  $\pi^+$  & 5 &  $0.65$  &  $0.53$  &  $0.57$  & $0.56$ \\
\noalign{\vspace{1pt}}
          &  $\pi^-$  & 5 &  $2.67$  &  $2.64$  &  $1.60$  & $1.62$ \\
\noalign{\vspace{1pt}}
  HERMES  &  $\pi^0$  & 5 &  $0.48$  &  $0.46$  &  $0.47$  & $0.43$ \\
\noalign{\vspace{1pt}}
          &  $K^+$    & 5 &  $6.14$  &  $5.71$  &  $3.87$  & $2.11$ \\
\noalign{\vspace{1pt}}
          &  $K^-$    & 5 &  $0.79$  &  $0.80$  &  $0.73$  & $0.69$ \\
\noalign{\vspace{5pt}}
\cline{1-7}
\noalign{\vspace{5pt}}
          &  $\pi^+$    & 9 &  $0.65$  &  $0.57$  &  $0.39$  & $0.64$ \\
\noalign{\vspace{1pt}}
  COMPASS &  $\pi^-$    & 9 &  $0.83$  &  $0.55$  &  $0.49$  & $0.70$ \\
\noalign{\vspace{1pt}}
          &  $K^+$      & 9 &  $0.84$  &  $0.75$  &  $0.78$  & $0.57$ \\
\noalign{\vspace{1pt}}
          &  $K^-$      & 9 &  $1.47$  &  $1.34$  &  $1.17$  & $1.36$ \\
\noalign{\vspace{3pt}}
\end{tabular}
\end{ruledtabular}
\label{rtab}
\end{table}
%
%
\begin{figure}
\begin{center}
\includegraphics[width=0.55\textwidth,bb= 10 150 580 660,angle=-90]
{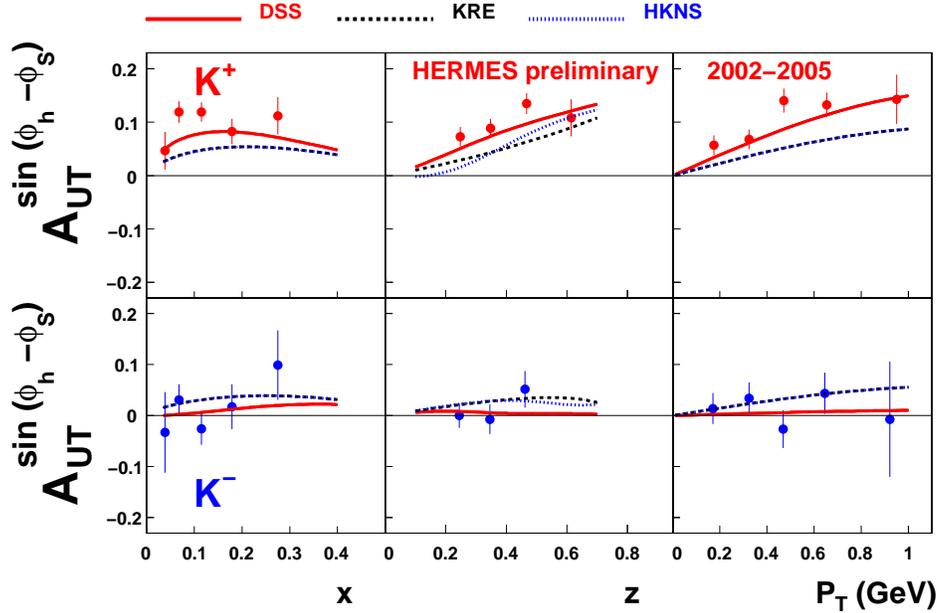} \hskip 2.85cm
\caption{\label{fig:dflo-kretz}
The results obtained from our fit using the kaon fragmentation functions
as given by de Florian {\it et al.} in Ref.~\cite{deFlorian:2007aj} (solid
lines) are compared with the results we would find by using the
KRE~\cite{Kretzer:2000yf}(dotted lines) and HKNS~\cite{Hirai:2007cx}
(dashed lines) sets of fragmentation functions.}
\end{center}
\end{figure}
%
\begin{figure}[t]
\includegraphics[width=0.7\textwidth,bb= 10 40 540 740,angle=0]
{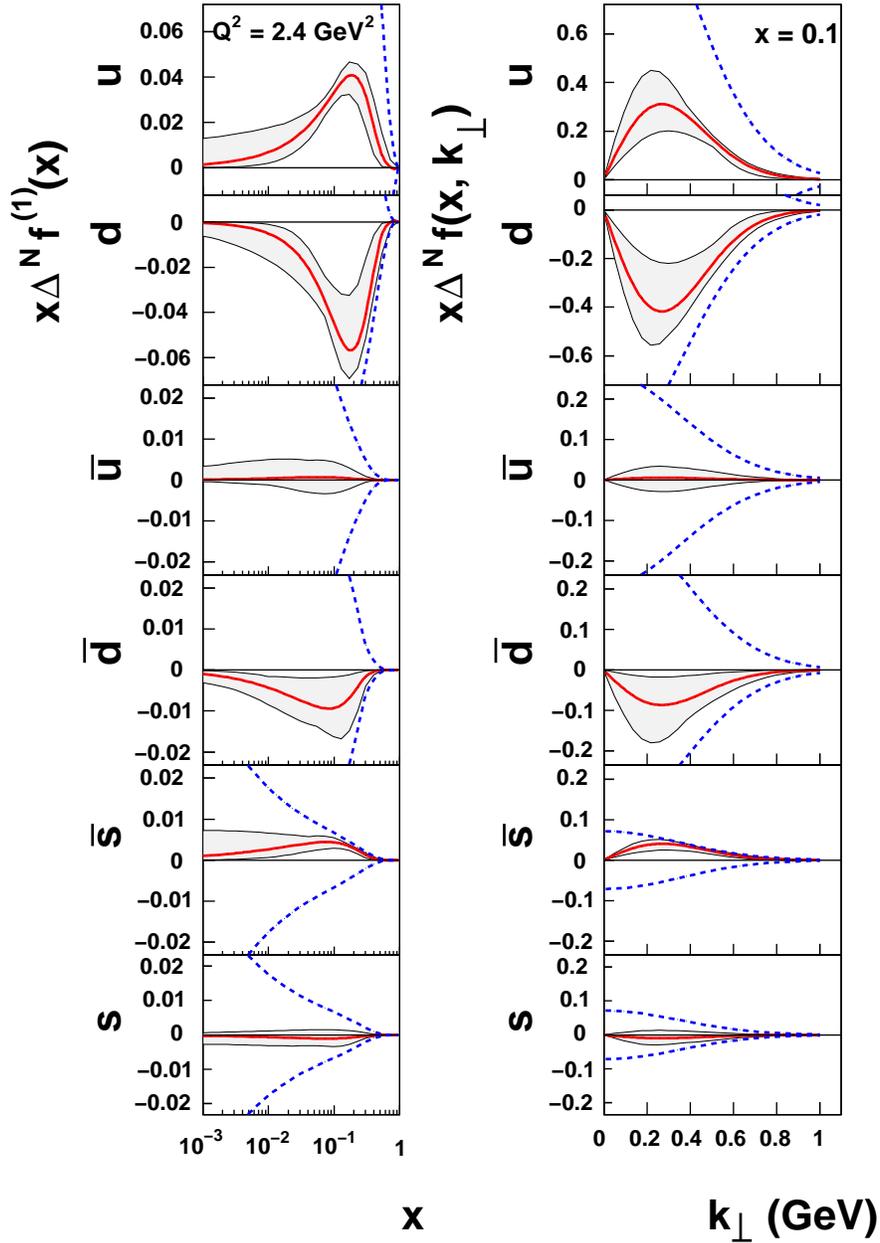}
\caption{\label{fig:sivers}
The Sivers distribution functions for $u$, $d$ and $s$ flavours, at the scale
$Q^2=2.4$ (GeV$/c)^2$, as determined
by our simultaneous fit of HERMES and COMPASS data (see text for details).
On the left panel, the first moment $x\,\Delta^N \! f^{(1)}(x)$,
Eq.~(\ref{mom}), is shown as a function of $x$ for each flavour, as indicated.
Similarly, on the right panel, the Sivers distribution
$x\,\Delta^N \! f(x,\kt)$
is shown as a function of $\kt$ at a fixed value of $x$ for each flavour, as
indicated. The highest and lowest dashed lines show the positivity limits
$|\Delta^N \! f| = 2f$.
}
\end{figure}
%
\begin{figure}[t]
\includegraphics[width=0.60\textwidth,bb= 10 140 580 660,angle=-90]
{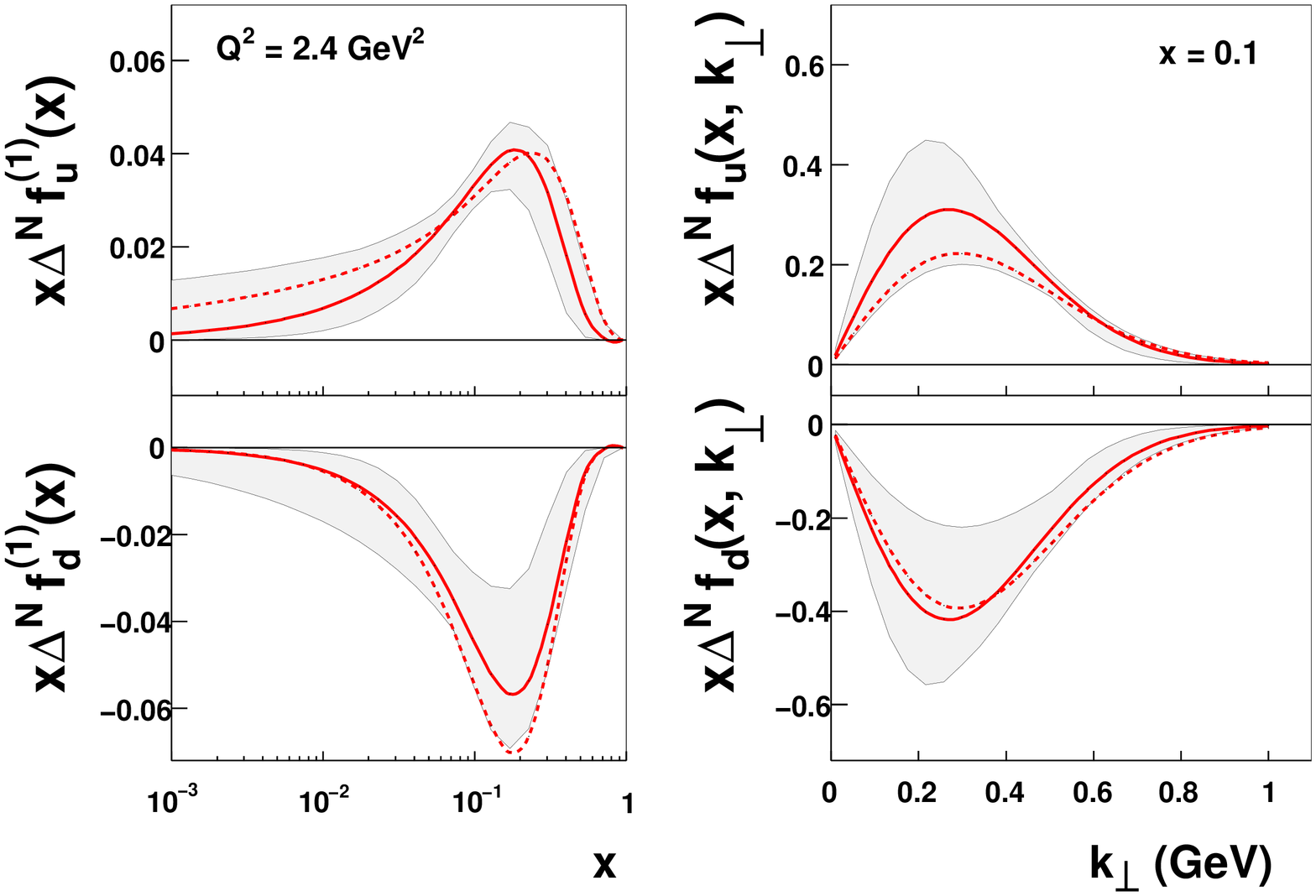}
\caption{\label{fig:sivers-como}
The Sivers distribution functions for $u$ and $d$ flavours, at the scale
$Q^2=2.4$ (GeV$/c)^2$, as determined
by our present fit (solid lines), are compared with those of our previous fit
\cite{Anselmino:2005ea} of SIDIS data (dashed lines), where $\pi^0$ and kaon
productions were not considered and only valence quark contributions were taken
into account. This plot clearly shows that the Sivers functions previously
found are consistent, within the statistical uncertainty bands, with the Sivers
functions presently obtained.
}
\end{figure}
In order to check the dependence on the set of unpolarized PDF's adopted,
we have also performed the fit by using the CTEQ6L~\cite{Pumplin:2002vw}
and the MRST01LO~\cite{Martin:2002dr} sets; in both cases,
the quality of the fit and the central-value results for the asymmetries are
so similar to those obtained with the GRV98LO set that they would be hardly
distinguishable in Figs.~\ref{fig:hermes} and~\ref{fig:compass}.

The shaded areas in Figs.~\ref{fig:hermes} and \ref{fig:compass}
(and in all subsequent figures where they are shown) represent statistical
uncertainties and correspond to a 95.45\% Confidence Level (CL): they are
determined according to the procedure described in Appendix~\ref{stat}.
Notice that further uncertainties of theoretical nature, intrinsic to
our phenomenological approach, are present and might widen the size of the
statistical bands. However, these are very difficult to assess: it suffices
to recall that our analysis is performed assuming a simple factorized $k_\perp$
dependence in Eqs.~(\ref{sivfac}), (\ref{partond}) and (\ref{partonf}), that
the actual $Q^2$ evolution of the Sivers function is unknown and that
uncertainties in the fragmentation functions have not been taken into account.
The functional form for the $x$-dependence of the Sivers functions used
in the fit, Eq.~(\ref{siversx}), is a simple one and more structured dependences
might allow better fits, with the statistical shaded areas covering better the
experimental errors of the data. At this stage, considering the available
experimental information and the remaining theoretical issues to be clarified,
we do not think that further refinements of our analysis would be meaningful.

Notice that in Fig.~\ref{fig:compass} we also show the results for $\pi^0$ at
COMPASS, for which no data is so far available, computed using our extracted
Sivers functions as given in Table~\ref{fitpar_sivers}. Similarly we have
computed $A_{UT}^{\sin(\phi_h-\phi_S)}$ for $K^0_S$ production at HERMES and
COMPASS and show them respectively in Figs.~\ref{fig:hermes} and \ref{fig:compass}.
As the $K^0_S$ is an equal mixture of $K^0 = d\bar s$ and $\bar K^0 = \bar ds$,
we have assumed isospin invariance, writing the $K^0_S$ FFs in terms of the
$K^+$ ones -- which are taken from Ref.~\cite{deFlorian:2007aj} -- as:
\bea
&& D_d^{K^0_S} = D_{\bar d}^{K^0_S} = \frac 12 \left[ D_u^{K^+}
+ D_{sea}^{K^+} \right] \nonumber \\
&& D_{\bar s}^{K^0_S} = D_s^{K^0_S} = \frac 12 \left[ D_{\bar s}^{K^+}
+ D_{sea}^{K^+} \right] \label{D0SFF} \\
&& D_{u}^{K^0_S} = D_{\bar u}^{K^0_S} = D_{sea}^{K^+} \equiv
D_{d}^{K^+} = D_{\bar u}^{K^+} = D_{s}^{K^+} = D_{\bar d}^{K^+} \,. \nonumber
\eea
Our computation of the $K^0_S$ asymmetry at COMPASS can be compared with the
available data~\cite{Alekseev:2008dn}, as shown in the upper-right plots of
Fig.~\ref{fig:compass}. Notice that these curves, contrary to the others in
the same figure, are not best fits, but a simple estimate, based on the
extracted Sivers functions and the adopted fragmentation functions of
Eq.~(\ref{D0SFF}).

In Fig.~\ref{fig:dflo-kretz}, our results, obtained using the kaon fragmentation
functions as given by de Florian {\it et al.} in Ref.~\cite{deFlorian:2007aj}
(solid lines), are compared with the best fit we would find by using the
KRE~\cite{Kretzer:2000yf} (dotted lines) and HKNS~\cite{Hirai:2007cx} (dashed
lines) sets of fragmentation functions. It is clear that the use of the new --
strange-quark sensitive -- fragmentation functions yields a much better
agreement with the experimental measurements of the SIDIS azimuthal
asymmetries for kaon production.

The Sivers functions generated by our best fit procedure are presented, at the scale $Q^2=2.4$ (GeV$/c)^2$, in
Fig.~\ref{fig:sivers}, where we plot, on the left panel, the first
$\bfk_\perp$ moment defined as
\be
\Delta^N \! f_{q/\pup}^{(1)}(x) \equiv \int d^2 \, \bfk_\perp \,\frac{\kt}{4m_p}
\, \Delta^N \! f_{q/\pup}(x, \kt) = - f_{1T}^{\perp (1) q}(x) \>, \label{mom}
\ee
and, on the right panel, the $\kt$ dependence of
$\Delta^N \! f_{q/\pup}$ at a fixed value of $x=0.1$. The highest and lowest
dashed lines show the positivity limits $|\Delta^N \! f| = 2f$.

Our results both confirm previous conclusions on the $u$ and $d$ Sivers
distributions and, despite the still large uncertainties
(see Table~\ref{fitpar_sivers}), offer some new clear information about
the so far unknown sea-quark Sivers functions. Let us comment in detail:
\begin{itemize}
\item
The HERMES data on kaon asymmetries, surprisingly large for $K^+$, cannot be
explained without a sea-quark Sivers distribution. In particular, we
definitely find
\be
\Delta^N \! f_{\bar s/\pup} > 0
\label{func_brokens}
\ee
and confirm the previous findings for valence flavours
\cite{Anselmino:2005ea, Vogelsang:2005cs, Collins:2005ie, Anselmino:2005an},
\be
\Delta^N \! f_{u/\pup} > 0 \quad\quad\quad  \Delta^N \! f_{d/\pup} < 0 \>.
\label{func_brokenv}
\ee
There are simple reasons for the above results. The Sivers
distribution function for $\bar s$ quarks turns out to be definitely positive,
due to the large positive value of $A_{UT}^{\sin(\phi_h-\phi_S)}$ for $K^+$;
notice that the value of $N_{\bar s}$ saturates the positivity bound
$|N_q| \leq 1$. Similarly, the positive sign of $\Delta^N \! f_{u/\pup}$ is,
essentially, driven by the positive $\pi^+$ and $K^+$ SSAs and the opposite sign of
$\Delta^N \! f_{d/\pup}$ by the small SSA measured by COMPASS on a deuteron
target. The $u$ and $d$ Sivers functions are also predicted to be opposite
in the large $N_c$ limit~\cite{Pobylitsa:2003ty} and in chiral models
\cite{Drago:2005gz}.

\item
The Sivers functions for $\bar u$, $\bar d$ and $s$ quarks, instead, turn out
to have much larger uncertainties; even the sign of the $\bar u$ and $s$
Sivers functions is not fixed by available data, while $\Delta^N \!
f_{\bar d/\pup}$ appears to be negative.
This could be consistent with a positive contribution from
$u$ quarks, necessary to explain the large $K^+$ asymmetry, which is
decreased,
for $\pi^+$, by a negative $\bar d$ contribution.
One might expect correlated Sivers functions for $s$
and $\bar s$ quarks: we have actually checked that choosing $\Delta^N \!
f_{s/\pup} = \pm \Delta^N \! f_{\bar s/\pup}$ slightly worsens the
$\chi^2_{dof}$ (from 1 up to about 1.1), but still leads to a reasonable fit.

\item
We notice that the Burkardt sum rule \cite{Burkardt:2004ur}
\be
\sum_a \int dx \, d^2 \bfk_\perp \, \bfk_\perp \, f_{a/\pup}(x, \bfk_\perp)
\equiv  \sum_a \, \langle \bfk_\perp^a \rangle = 0 \>, \label{bsr}
\ee
where, from Eqs. (\ref{sivnoi}) and (\ref{mom}),
\bea
\langle \bfk_\perp^a \rangle &=& \left[ \frac{\pi}{2}\int_0^1 \!\! dx
\int_0^\infty  \!\! d\kt \,\kt^2 \, \Delta^N \! f_{a/\pup}(x,\kt) \right]
(\bfS \times \hat{\bfP})  \nonumber \\ 
&=& m_p \int_0^1 \!\! dx  \, \Delta^N \! f_{q/\pup}^{(1)}(x)
\, (\bfS \times \hat{\bfP}) \equiv \langle \kt^a \rangle \,
(\bfS \times \hat{\bfP})\,, \label{bsr3}
\eea
is almost saturated by $u$ and $d$ quarks alone at $Q^2 = 2.4$ (GeV$/c)^2$:
\be
\langle \kt^u \rangle + \langle \kt^d \rangle =
-17^{+37}_{-55} \>\> ({\rm MeV}/c)\quad\quad\quad
\langle \kt^{\bar u} \rangle + \langle \kt^{\bar d} \rangle
+\langle \kt^s \rangle + \langle \kt^{\bar s} \rangle =
-14^{+43}_{-66} \>\> ({\rm MeV}/c). \label{bsr-u+d-sea}
\ee
The individual contributions for quarks are:
\bea
\langle \kt^u \rangle &=& 96^{+60}_{-28} \>\>
({\rm MeV}/c)\quad\quad\quad
\langle \kt^d \rangle \, = -113^{+45}_{-51} \>\>
({\rm MeV}/c)\nonumber \\
\langle \kt^{\bar u} \rangle &=& \>\> 2^{+24}_{-11} \>\>
({\rm MeV}/c)\quad\quad\quad
\langle \kt^{\bar d} \rangle \, = \,\,\> -28^{ +20}_{-60} \>\>
({\rm MeV}/c)\label{bsr-single} \\
\langle \kt^{s} \rangle &=& \!\! -4^{+11}_{-15} \>\>
({\rm MeV}/c)\quad\quad\quad
\langle \kt^{\bar s} \rangle \, = \quad\>\, 17^{+30}_{-8} \
({\rm MeV}/c)\>, \nonumber
\eea
thus leaving little room for a gluon Sivers function,
\be
-10 \le \langle \kt^g \rangle \le 48 \ ({\rm MeV}/c)\>, \label{bsr-g}
\ee
in agreement with other
similar results \cite{Anselmino:2006yq, Brodsky:2006ha}. The statistical uncertainties
in the values given above have been computed as explained at the end of
Appendix A.

\item
In Fig.~\ref{fig:sivers-como} we compare the $u$ and $d$ flavour Sivers
distribution functions, at the scale $Q^2=2.4$ (GeV$/c)^2$,
obtained in our present analysis with the $u$ and $d$
flavour Sivers functions we had found from our previous fit
\cite{Anselmino:2005ea}, where $\pi^0$ and kaon productions were not
considered, the Kretzer fragmentation function set was used, and only valence
quark contributions were taken into account in the polarized proton. This plot
shows that the Sivers functions previously obtained are consistent, within
the statistical uncertainty bands, with the Sivers functions presently found.
\end{itemize}

\section{Estimates for forthcoming experiments \label{estimates} }

\begin{figure}[t]
\includegraphics[width=0.35\textwidth,bb= 10 140 540 660,angle=-90]
{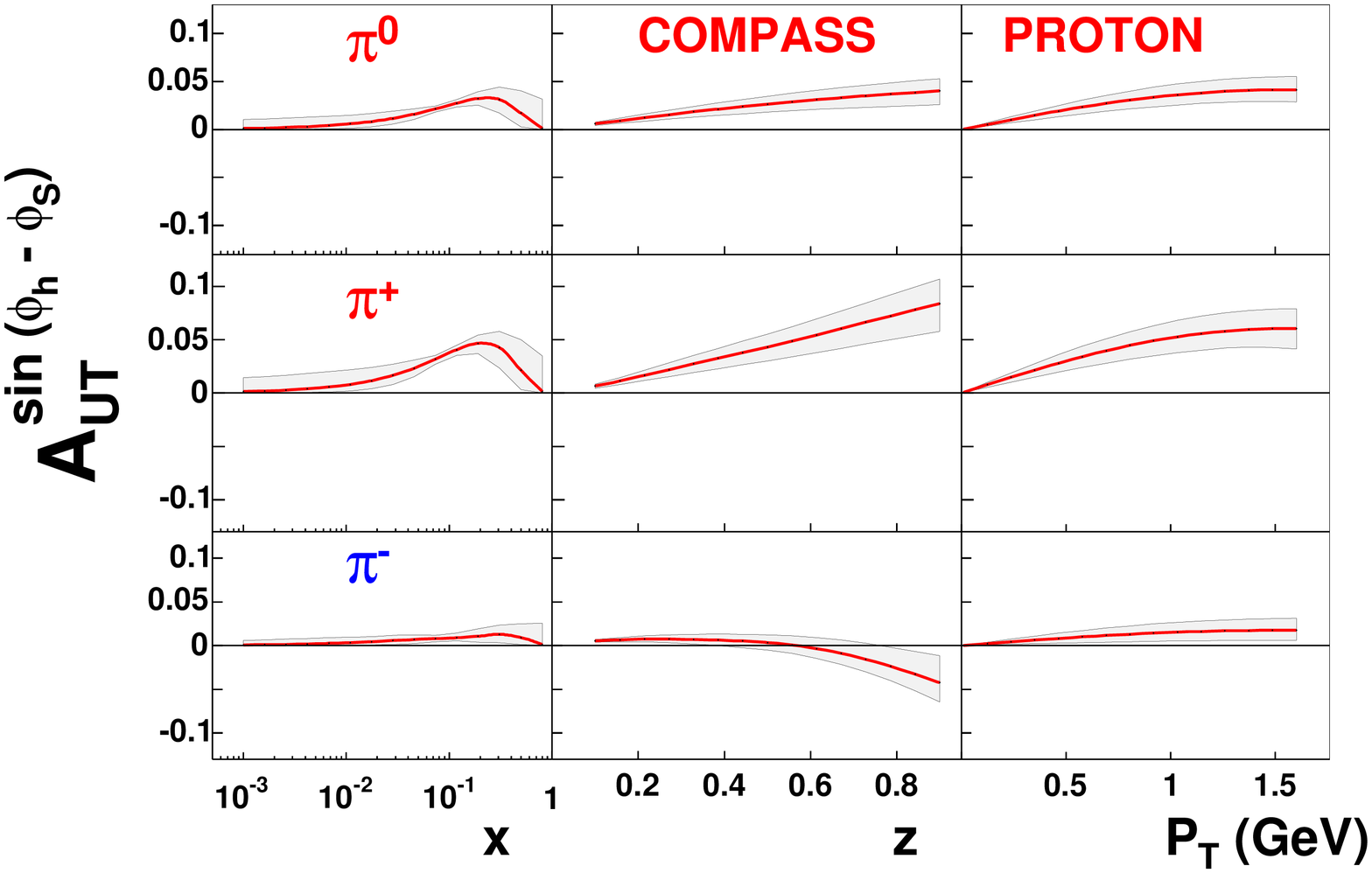} \hskip 2.85cm
\includegraphics[width=0.35\textwidth,bb= 10 140 540 660,angle=-90]
{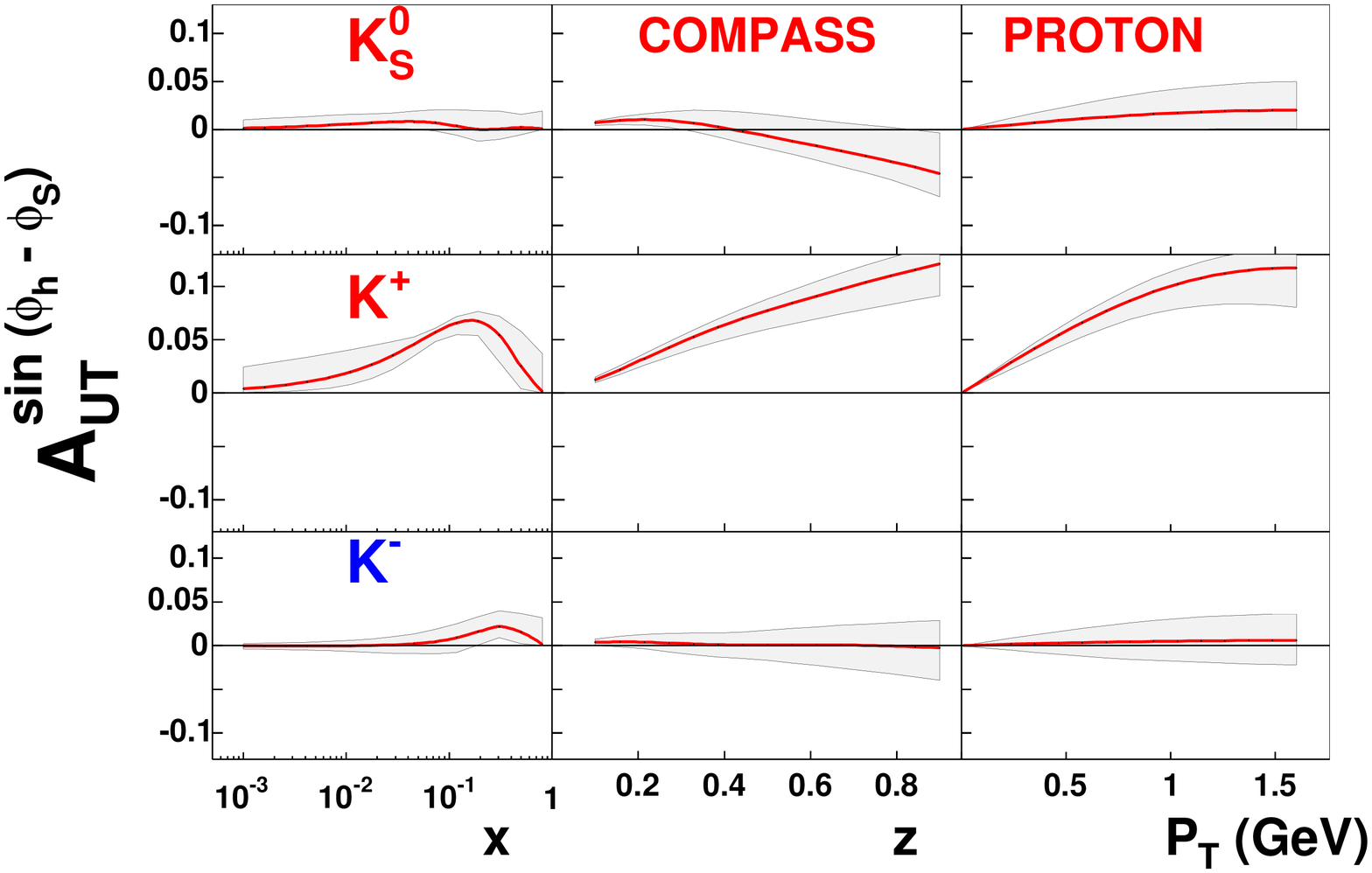}
\caption{\label{fig:compass_proton}
Estimates for the single spin asymmetry $A_{UT}^{\sin(\phi_h-\phi_S)}$
for pion and kaon production off a hydrogen target, which will be measured by
the COMPASS Collaboration.}
\end{figure}
%
\begin{figure}[t]
\includegraphics[width=0.35\textwidth,bb= 10 140 540 660,angle=-90]
{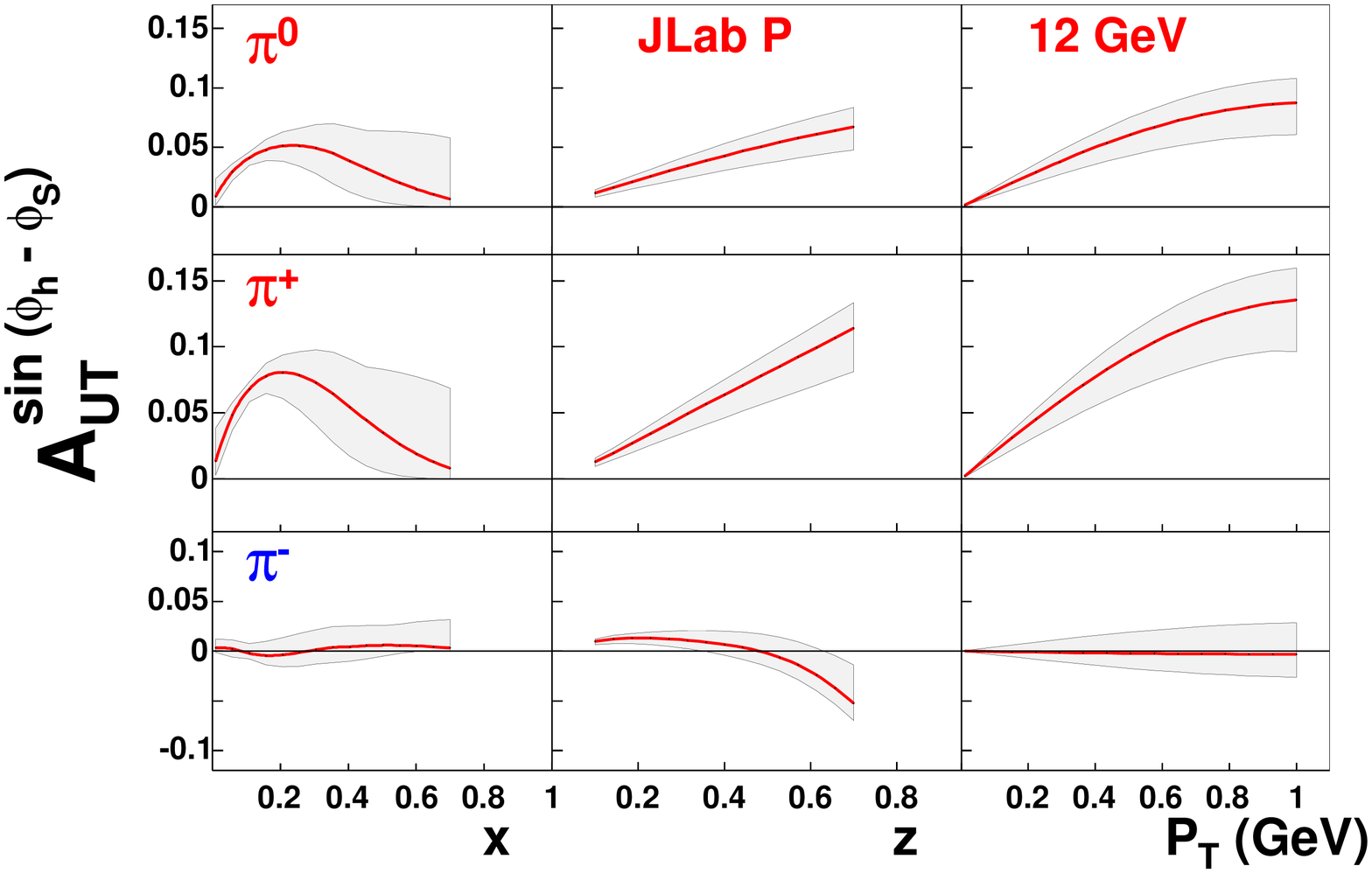} \hskip 2.7cm
\includegraphics[width=0.35\textwidth,bb= 10 140 540 660,angle=-90]
{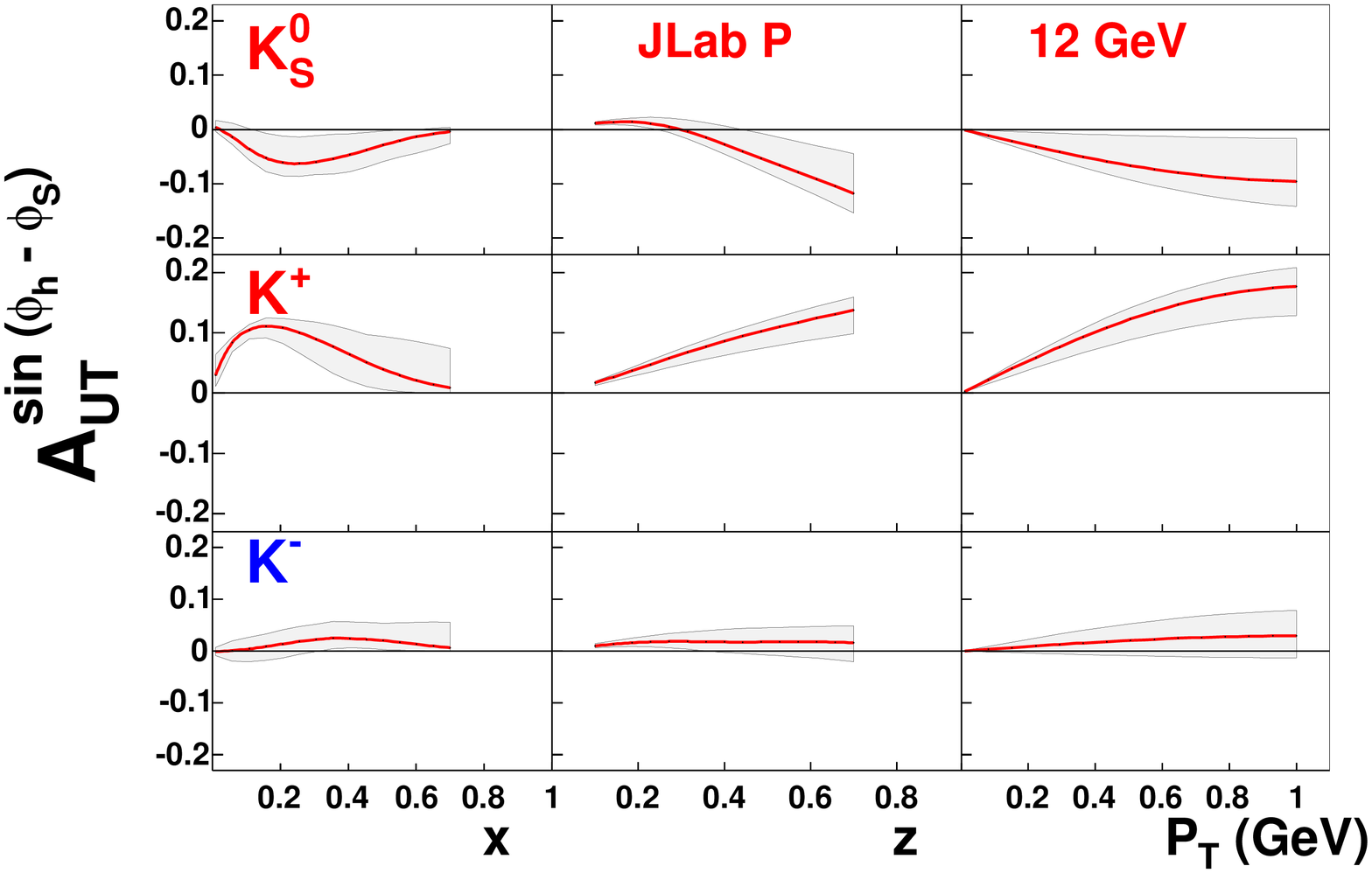}
\caption{\label{fig:jlab12_sivers-hyd}
Estimates for the single spin asymmetry $A_{UT}^{\sin(\phi_h-\phi_S)}$
for pion and kaon production, which will be measured at JLab operating on
a polarized hydrogen target with a beam energy of $12$ GeV.}
%
\includegraphics[width=0.35\textwidth,bb= 10 140 540 660,angle=-90]
{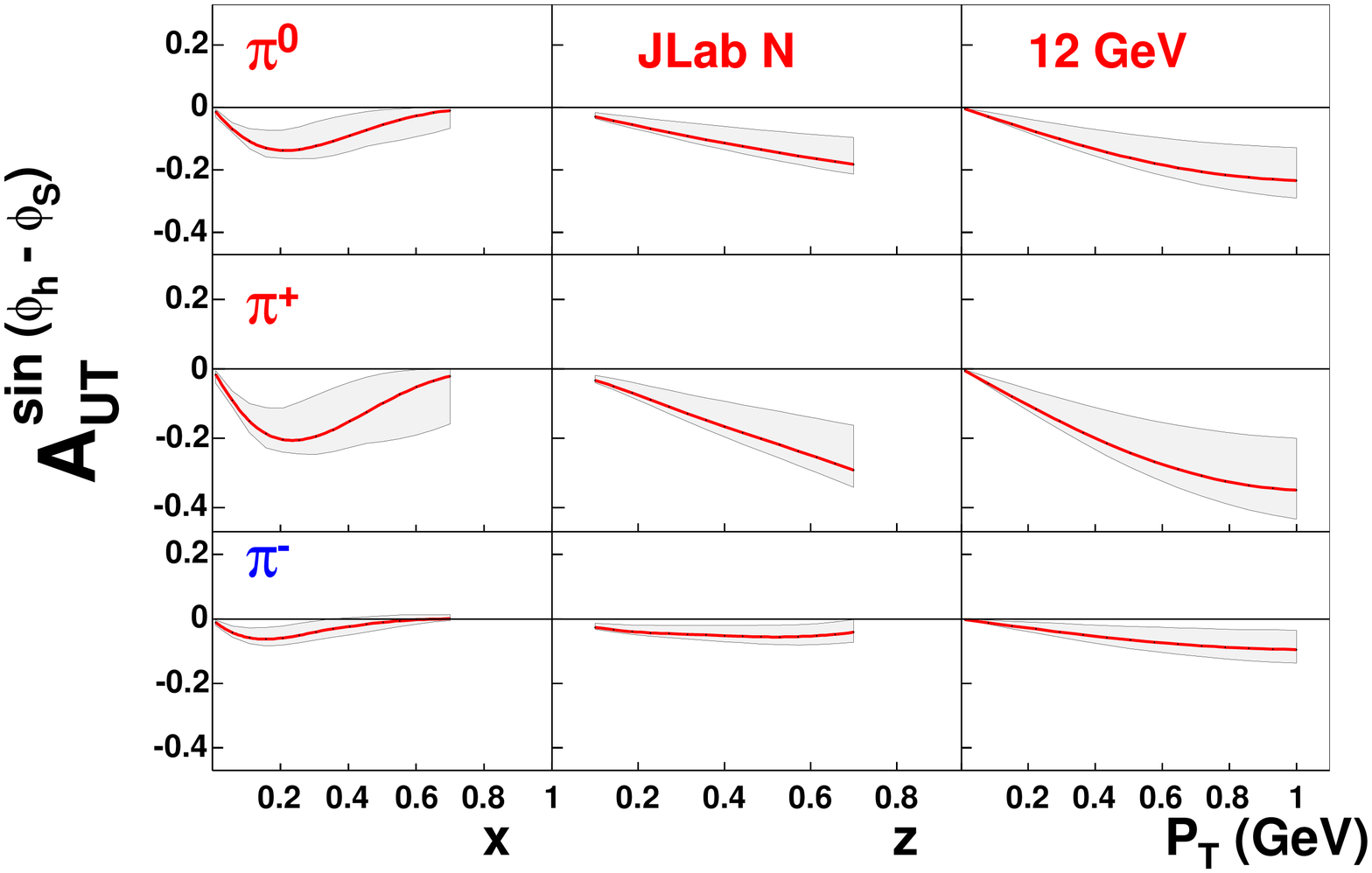} \hskip 2.7cm
\includegraphics[width=0.35\textwidth,bb= 10 140 540 660,angle=-90]
{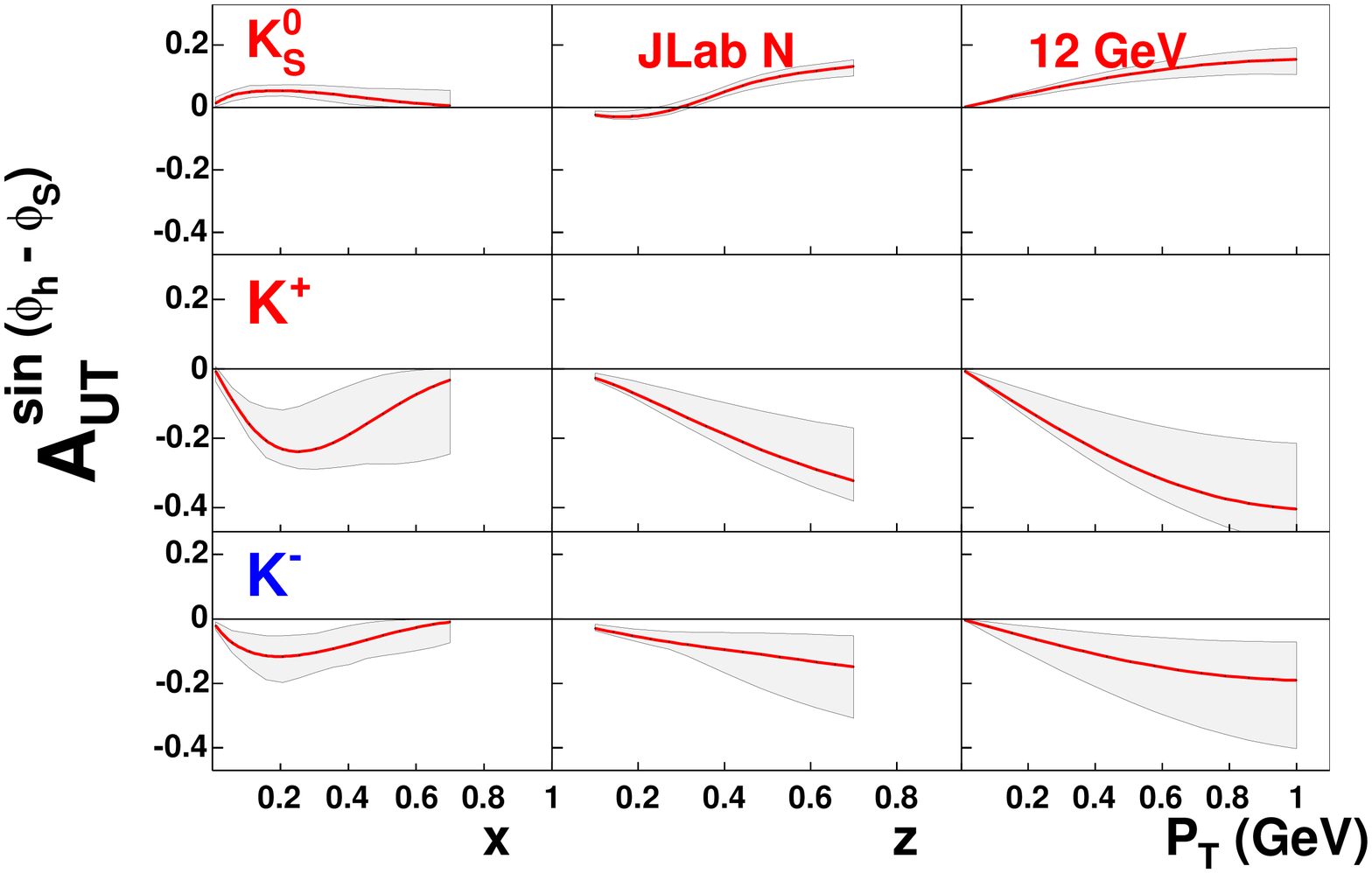}
\caption{\label{fig:jlab12_sivers-neu}
Estimates for the single spin asymmetry $A_{UT}^{\sin(\phi_h-\phi_S)}$
for pion and kaon production, which will be measured at JLab operating on
a polarized He$^3$ (neutron) target, with a beam energy of $12$ GeV.}
%
\includegraphics[width=0.35\textwidth,bb= 10 140 540 660,angle=-90]
{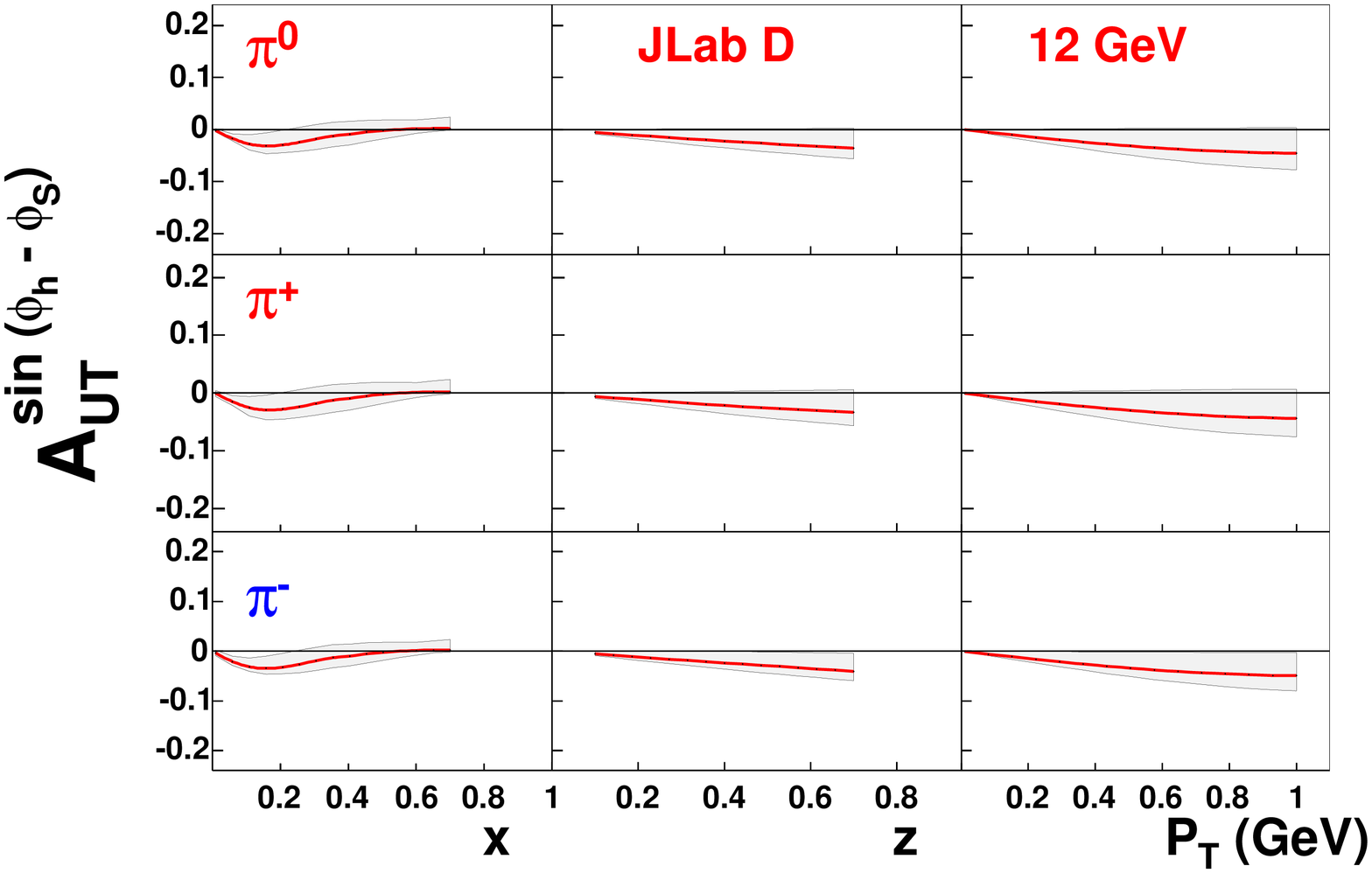} \hskip 2.7cm
\includegraphics[width=0.35\textwidth,bb= 10 140 540 660,angle=-90]
{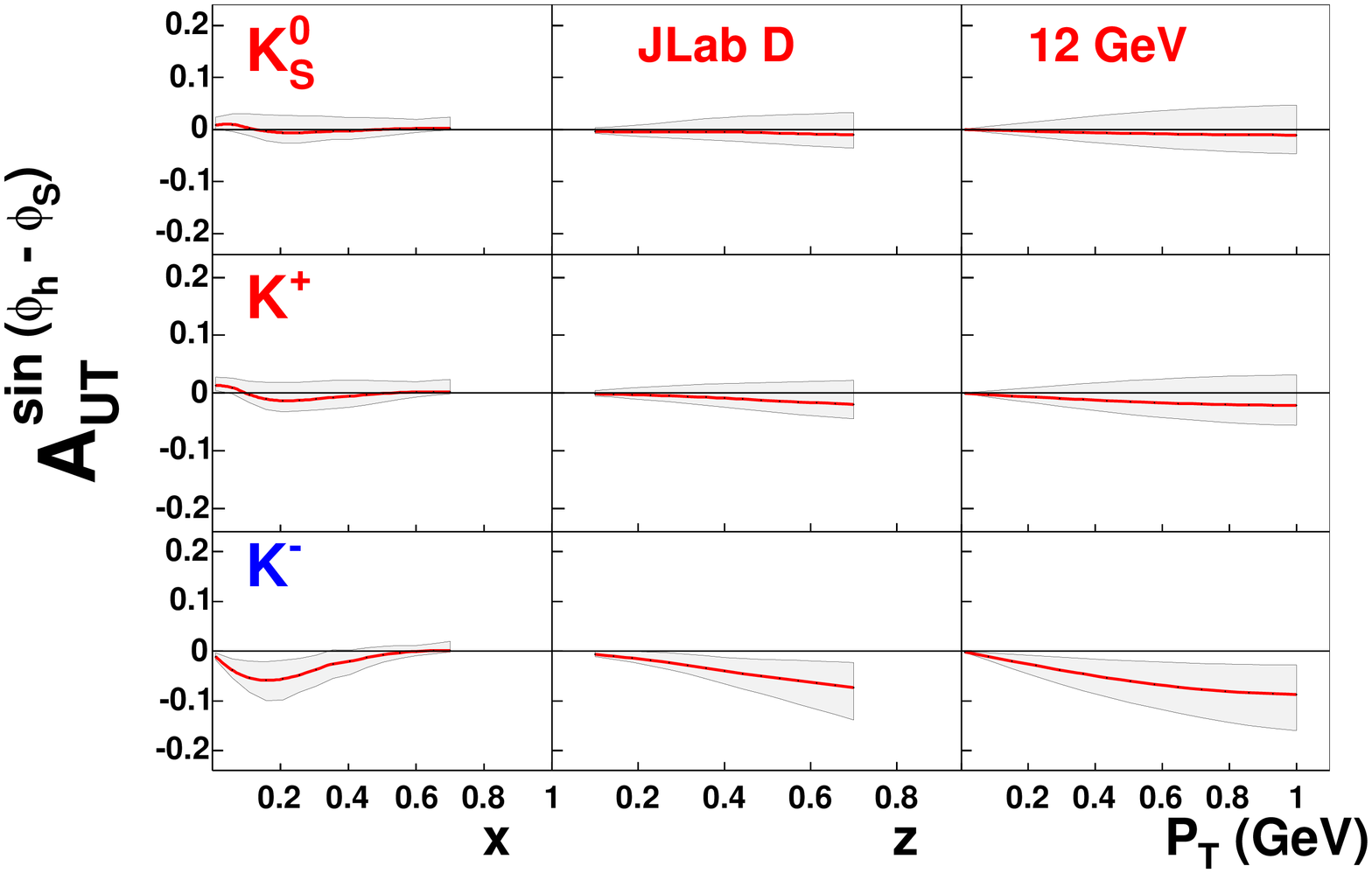}
\caption{\label{fig:jlab12_sivers-deu}
Estimates for the single spin asymmetry $A_{UT}^{\sin(\phi_h-\phi_S)}$
for pion and kaon production at JLab off a polarized deuteron target, with a beam energy
of $12$ GeV.}
\end{figure}

Using the Sivers functions determined through our fit, we can give
estimates for other transverse single spin asymmetries
$A_{UT}^{\sin(\phi_h-\phi_S)}$ which will be measured in the near future.
Fig.~\ref{fig:compass_proton} shows the results we obtain for the COMPASS
experiment operating with a hydrogen target, adopting the same experimental
cuts which were used for the deuterium target (Eq.~(71) of
Ref.~\cite{Anselmino:2005nn}).

Forthcoming measurements at the energies of $6$ and $12$ GeV are going to be
performed at JLab, on transversely polarized proton, neutron and deuteron
targets. The obtained data will be important for several reasons; they will
cover a kinematical region corresponding to large values of $x$, a region
which is so far unexplored by other SIDIS measurements. In particular,
a combined analysis of HERMES, COMPASS and JLab SIDIS data will allow a much
better determination of the $\beta$ parameters, which control the large $x$
behaviour of the Sivers distribution functions. In addition, the combined
analysis of proton and neutron target events will help flavour disentangling
and a more precise determination of $u$ and $d$ quark contributions.

Our estimates for the JLab SSAs, for pion and kaon production off proton,
neutron and deuteron targets, at 12 GeV, are presented in
Figs.~\ref{fig:jlab12_sivers-hyd}--\ref{fig:jlab12_sivers-deu}. At this
energy relatively large $Q^2$ values are available and the TMD factorization,
valid for $P_T\simeq \kt \simeq \Lambda _{\rm QCD} \ll Q$, should hold.
The adopted experimental cuts for a proton or a deuteron target are, in terms
of the usual SIDIS variables, the following:
\be
\begin{tabular}{lll}
$0.3 \le z_h \le 0.8\;\;\;$ ~~~~~~~~ & $0.05 \le P_T \le 1.5 \;{\rm GeV}/c$ \\
$0.05 \le \xb \le 0.7$  ~~~~~~~~ & $0.25 \le y \le 0.85$ \\
$1 \le Q^2 \le 8 \;({\rm GeV}/c)^2 $ ~~~~~~~~ & $W^2 \ge 4\;{\rm GeV}^2$ \\
$1.5 \le E_h \le 3.5\;  {\rm GeV}\;,$  & ~~~~~~~~
\end{tabular}
\label{JLab-12}
\ee
whereas for a neutron target they are:
\be
\begin{tabular}{lll}
$0.3 \le z_h \le 0.7$ ~~~~~~~~ &
$0.05 \le \xb \le 0.55$ \\
$0.34 \le y \le 0.9$ ~~~~~~~~ &
$Q^2 \ge 1 \;({\rm GeV}/c)^2$ \\
$W^2 \ge 2.3\; {\rm GeV}^2 \;,$  ~~~~~~~~ & ~
\end{tabular}
\label{JLab-neutron-12}
\ee
where, at order $(\kt/Q)$, one has $\xb = x$ and $z_h = z$; the exact
relationships can be found in Ref.~\cite{Anselmino:2005nn}.

Notice that these estimates, while well constrained by the available SIDIS
data at small $x$ values, might be less stringent at large $x$ values:
for example, relaxing the assumption of a unique $\beta$ value for all
flavours, would only marginally affect our present fit of HERMES and COMPASS
data, but would much widen the uncertainty band above $x \simeq 0.4$.

We have given complete sets of estimates for charged and neutral pions and
kaons. While the $\pi^\pm, \, \pi^0$ and $K^\pm$ computations originate from a
consistent procedure which uses fragmentation functions and Sivers distributions
obtained by fitting data involving the same particles, the $K^0_S$ estimates
might be affected by a greater uncertainty about its fragmentation functions,
which require the assumptions of Eq.~(\ref{D0SFF}). This can be seen in the
comparison between data and computations of Fig.~\ref{fig:compass}.

We have also computed estimates for JLab operating at 6 GeV, with the
corresponding kinematical cuts:
\be
\begin{tabular}{lll}
$0.4 \le z_h \le 0.7$ ~~~~~~~~ & $0.02 \le P_T \le 1 \;{\rm GeV}/c$\\
$0.1 \le \xb \le 0.6$ ~~~~~~~~ & $0.4 \le y \le 0.85$ \\
$Q^2 \ge 1 \;({\rm GeV}/c)^2$ ~~~~~~~~ & $ W^2 \ge 4\; {\rm GeV}^2$ \\
$1 \le E_h \le 4\; {\rm GeV} \;,$ ~~~~~~~~ & ~~~~~~~~\end{tabular}
\label{JLab-6}
\ee
for a proton or a deuteron target, and
\be
\begin{tabular}{lll}
$0.46 \le z_h \le 0.59$ ~~~~~~~~ &
$0.13 \le \xb \le 0.40$ \\
$0.68 \le y \le 0.86$ ~~~~~~~~ &
$1.3 \le Q^2 \le 3.1 \; ({\rm GeV}/c)^2 $ \\
$5.4 \le W^2 \le 9.3 \;  {\rm GeV}^2 $ ~~~~~~~~ &
$2.385 \le E_h \le 2.404\;  {\rm GeV} \;,$
\end{tabular}
\label{JLab-neutron-6}
\ee
for a neutron target. The SSAs result to be almost identical to those obtained
at 12 GeV; therefore, we do not show them explicitely.

A word of caution is necessary when discussing JLab SIDIS observables at
6 GeV; in the high-$x$, relatively low-$Q^2$ JLab kinematical regime,
target and identified hadron mass effects (in particular for kaons),
large $x$ resummations and higher-twist effects might be significant.
This implies potential theoretical problems in the analysis of such data.
Notice, moreover, that a much better statistics should be expected at 12 GeV.

\section{Conclusions}

We have performed a comprehensive analysis of SIDIS data on Sivers azimuthal
dependences, taking advantage of new and more precise experimental results.
Particularly challenging are the HERMES data on kaon asymmetries, showing an
unexpectedly large value of $A_{UT}^{\sin(\phi_h-\phi_S)}$ for $K^+$.
Our results confirm and improve previous extractions of the $u$ and $d$ Sivers
distributions and offer first insights into the sea contribution to
the Sivers effect.

It turns out that the data demand a non vanishing, and large, Sivers
distribution for $\bar s$ quarks, which, coupled to a new set of fragmentation
functions enhancing the role of strange quarks, appears to be the only way, at
present, to explain the $K^+$ data. The other sea quark ($\bar u, \bar d, s$)
contributions are, at this stage, less well determined, although they also
seem to be non vanishing.

Taken at face value, the extracted Sivers distributions indicate a saturation
of the Burkardt sum rule mainly due to $u$ and $d$ quarks alone, which carry
almost opposite transverse momentum. The sea quark contribution is altogether
rather small. This seems to rule out a contribution from gluons and, somehow,
points towards a picture of the proton structure with the parton orbital motion
restricted to valence quarks.

Finally, we have used our extracted Sivers distributions to compute estimates
for ongoing and planned new experiments at COMPASS and JLab.

\acknowledgements
We are grateful to Carlo Giunti for stimulating discussions on the statistical
analysis of our results.
We acknowledge support of the European Community - Research Infrastructure
Activity under the FP6 ``Structuring the European Research Area''
program (HadronPhysics, contract number RII3-CT-2004-506078).
M.A.,~M.B., and A.P.~acknowledge partial support by MIUR under Cofinanziamento
PRIN 2006.
This work is partially supported by the Helmholtz Association through
funds provided to the virtual institute ``Spin and strong QCD''(VH-VI-231).

\appendix

\section{$\chi^2$ analysis and statistical uncertainty bands \label{stat}}

For completeness, and because it is often a matter of debate, we briefly discuss
the techniques used for evaluating the statistical uncertainties of our estimates.
A standard (see Refs.~\cite{PDBook}, \cite{NR}) $\chi^2$ analysis is applied in
order to estimate the values of $M$ unknown parameters
${\bf a} = \{a_1, ..., a_M\}$. The total $\chi^2$ is calculated by
\bea
\chi^2 = \sum_{i = 1}^{N}\left(\frac{y_i - F(x_i; {\bf a})}{\sigma_i}\right)^2,
\label{chi2}
\eea
where we have a set of $N$ experimental measurements $y_i$ at known points
$x_i$. Each measurement is supposed to be Gaussian distributed with variance
$\sigma_i^2$.
The theoretical estimate $F(x_i; {\bf a})$ of the measurement $y_i$ depends
{\it non-linearly} on the $M$ unknown parameters $a_i$.

Minimizing the total $\chi^2$ yields a set of parameters ${\bf a}_0$ and a
value $\chi^2_{min}=\chi^2({\bf a}_0)$.

In our particular case we have $N = 173$ and, for the ``broken sea" ansatz, $M = 11$, thus
$n_{d.o.f.} = 162$; the minimum $\chi^2$ found by MINUIT is
$\chi^2_{min} = 162.65$.

At this stage, we would like to estimate the possible errors on the extracted
parameters and the statistical uncertainties on the corresponding Sivers functions and on
our estimates for the asymmetries which will be measured in future
experiments.

Let us call ${\bf a}_j$ some sets of parameters that could have been obtained
if a slightly different set of data $\{x_i,y_i\}_j$ were measured (within
experimental uncertainties one can {\it generate}, using Monte Carlo
techniques, new data sets and extract the corresponding parameter sets
${\bf a}_j$). Then, for each value of $j$, the quantity
\bea
\Delta \chi^2 \equiv \chi^2({\bf a}_j) - \chi^2({\bf a}_0)
\eea
is distributed according to a chi-square distribution, with
$M$ degrees of freedom (see sects.~32.3.2.3 of Ref.~\cite{PDBook}, 15.6 of
Ref.~\cite{NR}). In order to take into account the
correlations between all the parameters, we would like to perform a joint
estimation of $M$ parameters. The corresponding {\it coverage} probability
can be calculated according to the formula
\bea
\displaystyle P = \int_0^{\Delta \chi^2}\frac{1}{2\Gamma(M/2)}
\left(\frac{\chi^2}{2}\right)^{(M/2)-1}
\exp\left(-\frac{\chi^2}{2}\right){\rm d}\chi^2\, .
\label{P}
\eea
The meaning of the {\it coverage} probability can be explained as follows:
suppose that we generate sets of parameters ${\bf a}_j$, $j=1,..,I$,
which satisfy the condition
\bea
\chi^2({\bf a}_j) \le \chi^2_{min} + \Delta \chi^2\, ;
\eea
then, if the number of sets is large enough, we cover a hyper-volume in the
$M$ dimensional space which is called {\it confidence} region. The meaning of
the confidence region is that with a probability $P$ we will find the true set
of parameters  ${\bf a}_{true}$ inside this hyper-volume (see sect.~15.6 of
\cite{NR}). The minimal probability which is worth quoting is 68.3\% and is
historically connected to one sigma deviation of the normal distribution,
then 95.45\% corresponding to 2 sigma, etc.

Notice that if we wanted to estimate the error of one single parameter,
say $a_1$, having fixed all the other parameter values, then the previous
considerations would lead us back to the well known case in which $M=1$, and
$\Delta \chi^2 = 1$ for a required 68.3\% confidence level.

We determine the confidence hyper-volume corresponding to $P=0.9545$
coverage probability for the joint estimation of $M$ parameters. From
Eq.~(\ref{P}), we obtain
\bea
\Delta \chi^2_{fit} = 19.988\,.
\eea
We then generate 200 sets of parameters ${\bf a}_j$, $j=1,...,200$ which
satisfy the condition
\bea
\chi^2({\bf a}_j) \le \chi^2_{min} + \Delta \chi^2_{fit}\, ,
\eea
and cover our chosen confidence region.

Now, in all the plots
the central line corresponds to the extracted set of parameters ${\bf a}_0$
obtained from the $\chi^2_{min}$ value. In order to estimate the statistical
uncertainty
on this result we calculate the same quantity (either single spin asymmetry
or Sivers function) corresponding to the sets ${\bf a}_j$, $j=1,...,200$: at
each given point $x$ (or $z$ or $P_T$, as appropriate) the maximal and minimal
values among all of these give us the upper and lower uncertainty boundaries.
The resulting error band corresponds to the projection of the 95.45\%
confidence region onto a given observable. The meaning of this band is
straightforward: the probability to find the {\it true} result inside the
shaded corridor is 95.45\%.

\begin{figure}[t]
\includegraphics[width=0.37\textwidth,bb= 10 140 540 660,angle=-90]
{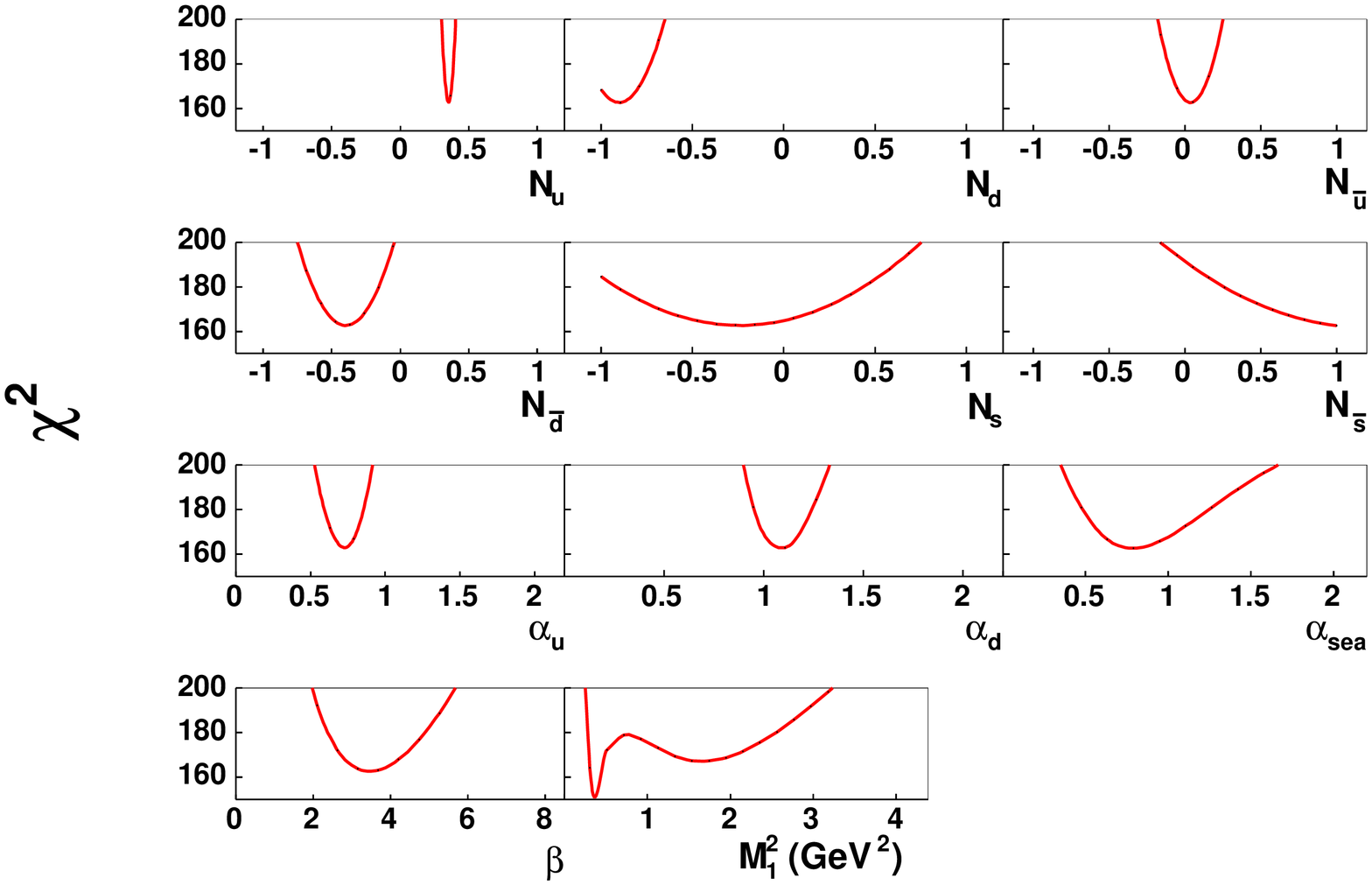} \hskip 1.85cm
\includegraphics[width=0.37\textwidth,bb= 10 140 540 660,angle=-90]
{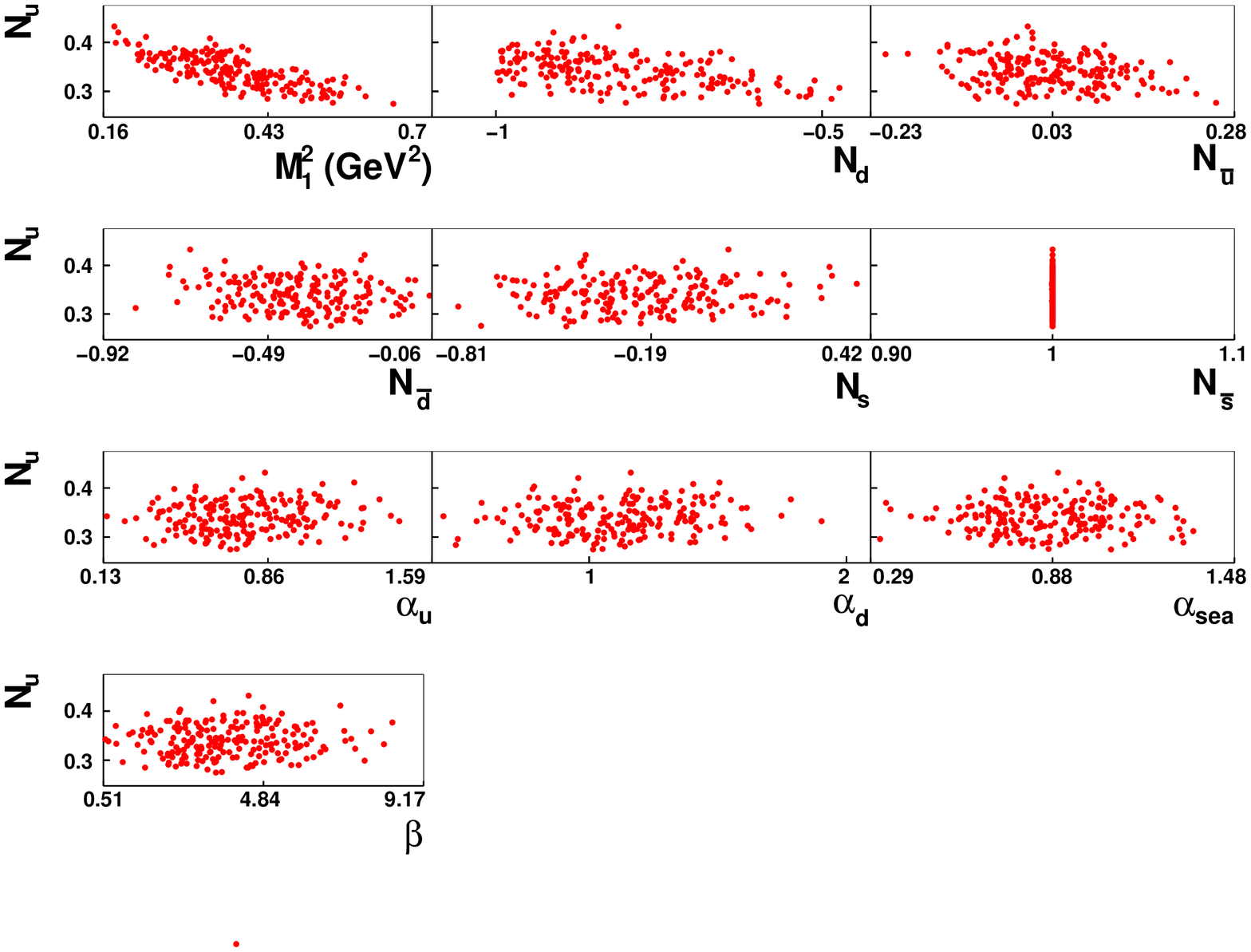}
\caption{\label{fig:stat}
The shape of $\chi^2$ as a function of the fit parameters (left) and the
scatter plot of the 200 generated sets as a function of $N_u$ (right).}
\end{figure}

The corresponding shape of the $\chi^2$ and the scatter plot of the 200
generated sets are shown in Fig.~\ref{fig:stat}, where one can clearly see
that the $\chi^2$ does not have a parabolic shape in the vicinity of
${\bf a}_0$; thus we need to take into account higher order corrections
to its Taylor expansion.

The explained method is general and is suitable to determine statistical uncertainties
in a general case, when the fitting function does not depend linearly on
parameters. Instead the ``standard'' method, exploiting the error matrix
issued by the MINUIT package, is applicable only in the cases when the
fitting function depends linearly on the parameters, and the parameter errors
are very small.

The statistical uncertainties in the values of $\langle \kt^q \rangle$ and their
combinations, Eqs.~(\ref{bsr-u+d-sea})-(\ref{bsr-g}), have been determined
consistently with the above procedure; these quantities have been computed for
each of the 200 sets of parameters and the upper and lower limits shown in
Eqs.~(\ref{bsr-u+d-sea})-(\ref{bsr-g}) simply correspond to the highest and
lowest values found.

\end{document}